\documentclass[fleqn,usenatbib]{mnras}

\usepackage{newtxtext,newtxmath}

\usepackage[T1]{fontenc}

\DeclareRobustCommand{\VAN}[3]{#2}
\let\VANthebibliography\thebibliography
\def\thebibliography{\DeclareRobustCommand{\VAN}[3]{##3}\VANthebibliography}

\usepackage{graphicx} 
\usepackage{amsmath} 

\newcommand{\jktebop}{\textsc{jktebop}}
\newcommand{\phoebe}{\textsc{phoebe2}}

\newcommand{\vcep}{V446~Cep}
\newcommand{\spd}{\textsc{spd}}
\newcommand{\bcep}{$\beta$~Cep}

\title[V446 Cep is a \bcep\ star in a multiple system]{V446 Cephei: a \bcep\ pulsator in a multiple system}

\author[A. Moharana et al.]
{A. Moharana,$^{1}$\thanks{E-mail: a.moharana@keele.ac.uk,  dr.ayush.moharana@gmail.com}
J. Southworth,$^{1}$ 
K. Pavlovski,$^{2}$ 
A. Miszuda,$^{3}$
R. S. Rathour,$^{3,8}$
K. G. He{\l}miniak,$^{4}$
\newauthor
F. Marcadon,$^{3}$
D. M. Bowman,$^{5,6}$
T. B. Pawar,$^{7}$
and 
A. Tkachenko$^{6}$
\\
$^{1}$Astrophysics group, Keele University, ST5 5BG, Staffordshire, UK\\
$^{2}$Department of Physics, Faculty of Science, University of Zagreb, 10 000 Zagreb, Croatia\\
$^{3}$Nicolaus Copernicus Astronomical Center, Polish Academy of Sciences, ul. Bartycka 18, 00-716 Warszawa, Poland\\
$^{4}$Nicolaus Copernicus Astronomical Center, Polish Academy of Sciences, ul. Rabia\'{n}ska 8, 87-100 Toru\'{n}, Poland\\
$^{5}$School of Mathematics, Statistics and Physics, Newcastle University, Newcastle upon Tyne NE1 7RU, UK \\
$^{6}$Institute of Astronomy, KU Leuven, Celestijnenlaan 200D, 3001 Leuven, Belgium\\
$^{7}$Villanova University, Dept.\ of Astrophysics and Planetary Sciences, 800 East Lancaster Avenue, Villanova, PA 19085, USA\\
$^{8}$Universit\'e Côte d'Azur, Observatoire de la C\^ote d'Azur, CNRS, Laboratoire Lagrange, France\\
}

\date{Accepted XXX. Received YYY; in original form ZZZ}

\pubyear{\the\year{}}

\begin{document}
\label{firstpage}
\pagerange{\pageref{firstpage}--\pageref{lastpage}}
\maketitle

\begin{abstract}
\bcep{}  stars in eclipsing binary (EB) systems give us an opportunity to put observational constraints on their structure and stellar parameters. We present a comprehensive analysis of the \bcep{} star in the EB \vcep\,  using \textit{TESS} photometry and HERMES spectra. We calculate the stellar and orbital parameters using light curve modelling and spectral disentangling. The EB has an orbital period of $3.808567 \pm 0.000012$ d and a mass ratio of $0.1550 \pm 0.0012$. We find the \bcep{} star to have a mass of $10.68 \pm 0.06$~$\mathrm{M_{\odot}}$,  a radius of $5.864 \pm 0.033$ $\mathrm{R}_{\odot}$, and a $T_{\rm eff}$ of $24220 \pm 180$~K. The secondary has a mass of $1.657 \pm 0.017$~$\mathrm{M_{\odot}}$,  a radius of $1.530 \pm 0.014$~$\mathrm{R}_{\odot}$, and a $T_{\rm eff}$ of $9080 \pm 390$~K. We also extract the abundances of C, N, O, Mg, and Si for the \bcep{} star, which are found to be consistent with galactic OB binaries. We identified 21 distinct pulsation frequencies, with the dominant mode at 10.24324 d$^{-1}$, which corresponds to a near-harmonic of the system's orbital frequency. The two stars in the EB have asynchronous rotation, with both stars rotating faster than the orbital frequency. We detect a companion to the EB using eclipse timing variations and period changes of the dominant pulsation frequency. We calculate the minimum mass of this tertiary companion to be $4.11 \pm 0.32$~$\mathrm{M_{\odot}}$ which is on an orbit of 2303$\pm$69~d around the EB. Using spectral energy distributions and MIST isochrones, we conclude that \vcep\ is either a co-evolving hierarchical 2+2 quadruple or a triple system where the third body is a compact object.

\end{abstract}




\section{Introduction}
Massive stars, even though they have a short lifetime and are limited in number, are involved in many important phenomena in the universe. Massive stars help in chemical enrichment 
as they produce heavy elements in the universe \citep{Maeder_1981,Wu_heavyelem_2021,Higgins_2023}. They also drive star formation via different feedback mechanisms \citep{Grudic_2019}. Massive stars are bright and can be observed in other galaxies. This makes them an important tool for calibrating the distance scale \citep{Pierce_RSG_2000} as well. They are progenitors of supernovae, Wolf-Rayet stars, and sources of gravitational wave mergers, among many other products of stellar evolution.  However, there remain issues in the theory of the evolution of massive stars that need improvement to fully understand the final products.

Current stellar evolution models for massive stars contains large theoretical uncertainties that are evident even in the main sequence phase. Since massive stars have radiative envelopes and convective cores, understanding the chemical mixing at the boundary between convective and radiative zones is important. Understanding chemical mixing also allows to probe changes main sequence lifetime and helium core masses, which propagates into compact remnant masses \citep{Schootemeijer_2019,Johnston_2021}. Rotation in massive stars also impacts the evolution by inducing mixing, and mass loss \citep{maeder2000,langer2012,2025MNRAS.543.2796H}. Magnetic fields add another layer of complexity to the problem of the evolution of massive stars \citep{Keszthelyi_2019,Keszthelyi_2024}.

To tackle such problems, one needs many observables, like radii, masses, temperatures, and composition, to break the intrinsic parameter degeneracies arising out of all the above processes. 

Asteroseismology is an avenue to obtain observables for understanding the stellar interior.
Asteroseismology with space-based photometry has improved our knowledge about stellar structure and the evolution of stars across the HR~diagram (see reviews by \citealt{Aerts2021} and \citealt{Kurtz2022}), from solar-type and red giant stars \citep{Bedding2011Nat,Huber2011,Miglio2012,ChaplinMiglio2013} to massive stars \citep{Bowman2020, Burssens2023}.  There are three main classes of  OB-type pulsators: \bcep{} stars, slowly pulsating B (SPB) stars, and stochastic low-frequency (SLF) variables (see \citealt{Bowman2023}). While SPB stars are of somewhat lower masses ($3 \lesssim M \lesssim 9$~$\mathrm{M_{\odot}}$), \bcep{} stars allow us to probe stars with masses above about 8~$\mathrm{M_{\odot}}$ and even upwards of 30~$\mathrm{M_{\odot}}$ \citep{Aerts_2010, Bowman2020}. Stars with SLF variability are found over a range of stellar masses and may be explained by various mechanisms, different from that of SPB and \bcep\, stars \citep{Bowman2023}. \bcep{} stars pulsate in pressure and gravity modes which are excited by the $\kappa$ mechanism within the Z-bump created by iron-peak elements \citep{dziembowksi_pam}. The pulsation periods typically range from 2 to 8~h \citep{Aerts_2010, Bowman2020}. These high-amplitude pulsations have been found in massive stars across a wide range of masses and ages \citep{Burssens2020, Fritzewski_2025} and provide us a diverse sample to study massive stars. 

Massive stars mostly exist in multiple systems \citep{Sana_2012, DucheneKraus2013,Offner2023}. This further complicates the analysis of massive stars by adding complex phenomena such as mass transfer and tidal effects \citep{Podsiadlowski_2010,deMink_2014,MarchantBodensteiner2024}. Multiplicity also affects stellar pulsations, especially in close binaries \citep{Guo_2021,SouthworthBowmanARAAreview}. For example, mass transfer can change pulsation periods of both pressure and gravity modes \citep{Wagg_2024,Miszuda_2025}. Tidal forces can perturb the axis of pulsation during the course of the orbital motion \citep{Bowman_2019_UGru,Steindl2021,vanReeth_2023}, and any also tilt the pulsation axis \citep{Fuller_2020,Handler_PTA,Jayaraman_2022}. In specific cases, the tides from pulsations, known as inverse tides, can even influence the orbital motion \citep{Fuller_2021}. On the other hand, we can also use multiplicity to our advantage and increase the number and type of observables to constrain the uncertainties in our understanding of massive stars.

Double-lined eclipsing binaries (DLEBs) provide us with among the most precise direct measurements of stellar masses and radii \citep{Torres_2010}. A pulsating star in a DLEB gives us the opportunity to connect the stellar structure with global stellar parameters. These systems have been documented in the literature since the 1970s (AB Cas; \citealt{Tempesti_1971}) but have been revolutionised because of space photometry --- see review by \citet{SouthworthBowmanARAAreview}.
Early space-based missions like \textit{MOST} \citep{Walker2003}, \textit{CoRoT} \citep{Baglin2006}, and \textit{Kepler} \citep{Borucki2010,Howell2014} were highly successful for asteroseismology of low-mass stars but generally lacked in observations of \bcep{} stars in EBs. While the \textit{BRITE} mission \citep{Weiss2014, Weiss2021} did observe some bright \bcep\, stars, the data quality was not as high as larger space missions. With the \textit{Transiting Exoplanet Survey Satellite} (\textit{TESS}; \citealt{Ricker_2015}), the scene is changing with new discoveries of \bcep\ in EBs \citep{HighMassPul_SouthBowm2022,EzeHandler_2024}. 
Still, to date there are only five confirmed \bcep{} stars in EBs with precise mass and radius estimates\footnote{DEBCat: \url{https://www.astro.keele.ac.uk/jkt/debcat/}} \citep{Debcat}. 

In this paper, we present a comprehensive analysis of \vcep, an EB with a \bcep\, component. \vcep\ (RA=22:08:45.59, $\delta$=61:01:20.7) is a bright spectroscopic binary in the Alessi-Teutsch~5 open cluster. It was first discovered as an EB by \cite{Kazarovets_1999} and its pulsating nature was first explored by \cite{HighMassPul_SouthBowm2022}. It was later analysed as a spectroscopic binary by \cite{Tkachenko_2024}, and is reported in the \textit{TESS} catalogue of \bcep{} stars in EBs by \citep{EzeHandler_2024}. 
We used ten sectors of \textit{TESS} observations together with high-resolution spectroscopy to obtain precise stellar parameters of the EB, as detailed in \autoref{sec:stellarparam}. A comprehensive study of the pulsations is presented in \autoref{sec:pulsations}. We also found the presence of a stellar companion to the EB, which is described in \autoref{sec:thirdbody}. Finally, we discuss the possible scenarios of formation and evolution of \vcep\, in \autoref{sec:evo}.

\section{Observations}
\subsection{Spectroscopy}

The spectroscopic observations included high-resolution \'{e}chelle spectra assembled with the high-efficiency
and high-resolution Mercator \'{e}chelle spectrograph (HERMES; \citealt{Raskin_2011}) mounted on the Mercator telescope at the Roque de los Muchachos Observatory on La Palma, Canary Islands, Spain. The HERMES spectrograph provides a high resolving power ($R \simeq 85\,000$ at 550~nm), and a large wavelength coverage from about 380 to 1000~nm.

 In total, we use 72 HERMES spectra of \vcep. The reduction and rectification of these spectra prior to our work were described in detail in \citet{Tkachenko_2024}. The final spectra from \citet{Tkachenko_2024} underwent additional detailed inspection and normalisation using custom routines prior to their use in this work.

\subsection{Photometry}

 The \textit{TESS} mission was launched in April 2018 and placed on an elliptical 13.70 d orbit around Earth. {\it TESS} has four wide-field CCD cameras which have a combined field of view of 24 $\times$ 96 deg$^2$. {\it TESS} observes a field (a sector of the whole sky) for about 27 d before changing the field. Over the years, some of the fields have been revisited by {\it TESS}, providing us with a long time base for the photometry of these fields.

\begin{figure}
    \centering
    \includegraphics[width=\columnwidth]{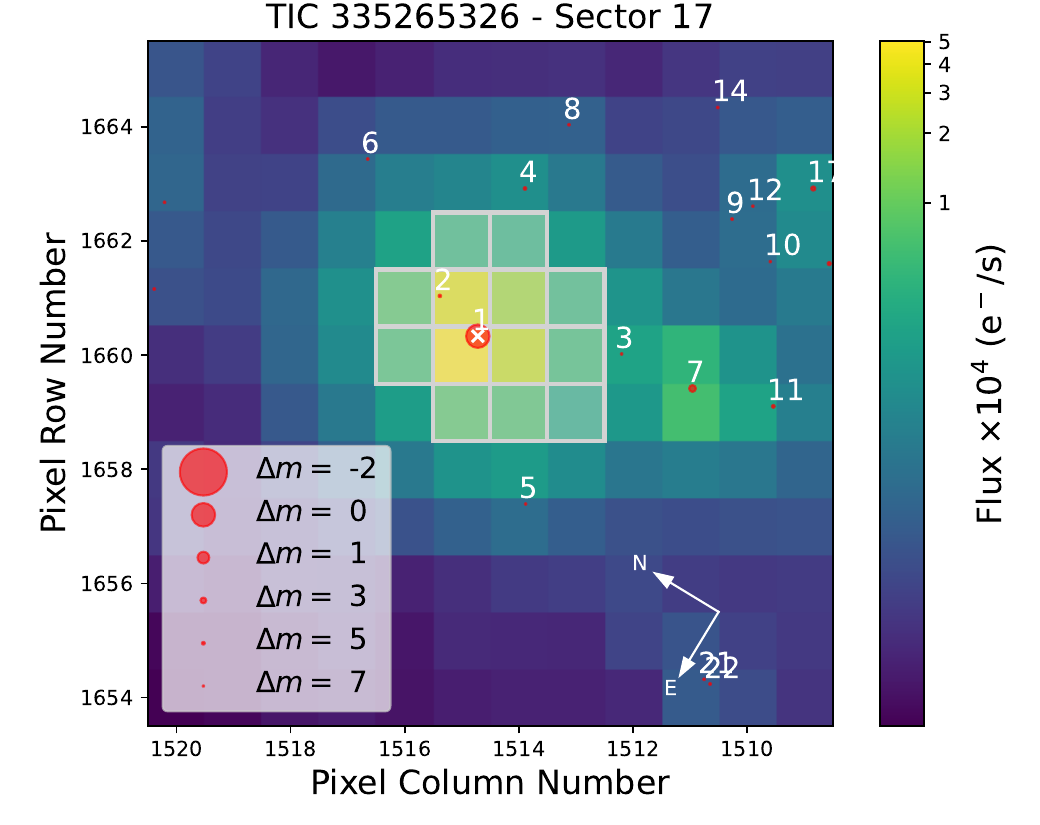}
    \caption{ \textit{TESS} TPF of \vcep\, with nearby sources. The marker sizes indicate the magnitude difference between \vcep\, and the respective star. Star 2 has a \textit{G}-band magnitude of 13 compared to 7.8 of V446 Cep. The white lines mark the pixels used in the SPOC aperture.}
    \label{fig:tpf}
\end{figure}

\vcep\ (TIC 335265326) was observed across ten TESS sectors\footnote{GI programmes: G022062, G06037, G06057, G05036, G05003},  specifically sectors 16, 17, 24, 56, 57, 76, 77, 83, 84, and 85. To check for possible contamination in the target pixel files (TPF) we used the code \textsc{tpfplotter}\footnote{\url{https://github.com/jlillo/tpfplotter}} \citep{tpfplotter_2020}. {\sc tpfplotter} overlays nearby stars from the \textit{Gaia} \citep{gaia2016} on the TPF. The photometric aperture used by the {\it TESS} Science Processing Operations Center (SPOC;  \citealt{TESS_SPOC}) is also available to plot on the TPF as shown in Fig.\ref{fig:tpf}.   We found that \vcep\ was the brightest star in the field. The only other star in the SPOC aperture  was a star with a {\it G}-band brightness of $\sim$ 13~mag, compared to 7.8~mag for \vcep{}. Therefore we decided to use the simple aperture photometry (SAP) light curves with 2 min cadence, which are produced by SPOC. 

We extracted the light curves from all sectors using \textsc{lightkurve}\footnote{\url{https://lightkurve.github.io/lightkurve/}} \citep{code_lightkurve}. Long-term and short-term trends in the light curves were cleaned subsequently using the \textsc{wotan}\footnote{\url{https://github.com/hippke/wotan}} package \citep{Hippke_2019}. The cleaning used the biweight time-windowed slider method \citep{tukeybiweight}. It involved two successive iterations with windows of 30~d and 8~d, respectively. We further removed parts of the light curve that did not have consistent eclipse depths by manual inspection. These were mostly incomplete eclipses near the gaps at the midpoint of each sector. This gave us the final light curve that we used for further analysis. 

\section{Stellar parameters}
\label{sec:stellarparam}
We extracted the stellar, orbital, and atmospheric parameters of the binary system using a combination of photometric and spectroscopic analyses which are described in the following subsections.

\subsection{Light curve modelling}
\label{sec:lightcurvemod}

\subsubsection{\jktebop}
We used version 40 of the code \jktebop{} \citep{jktebop} for initial modelling. We individually modelled light curves from every sector. 

\begin{figure*}
    \includegraphics[width=\textwidth]{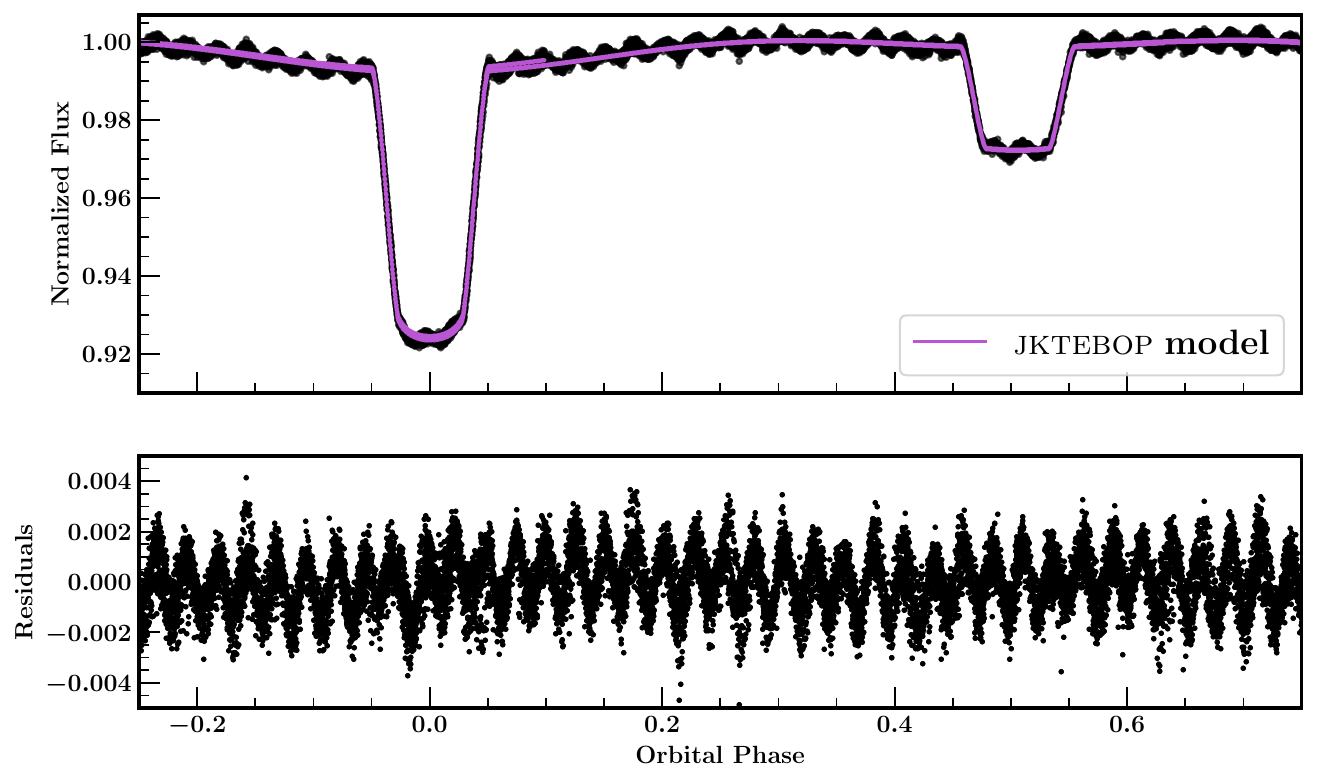}
    \caption{The top panel shows the binary phase-folded light curve and the binary model from \jktebop. The lower panel shows the residuals after subtraction of the binary model. The dominant pulsation ($\sim 10 \, \mathrm{d}^{-1}$) is clearly visible and has around 39 cycles in one orbital phase.}
    \label{fig:phasedjklc}
\end{figure*}

We initialised our models with parameters from \cite{HighMassPul_SouthBowm2022} and an approximate estimate of the time of primary minimum ($T_0$) from the first primary eclipse of each sector.  The initial set of free parameters included $T_0$, surface brightness ratio ($J$), sum of relative\footnote{defined relative to the semi-major axis ($a_1$)} radii ($r_{\rm 1}+r_{\rm 2}$), radius ratio ($k$), eccentricity ($e_1$) and argument of periastron ($\omega_1$) parametrised as $e_1\cos{\omega_1}$ and $e_1\sin{\omega_1}$, orbital period ($P_1$), and light scale factor ($L_0$). We also optimised the fraction of reflected light on each star, which parametrised the reflection effect. After the first round of optimisation, we calculated the residuals after subtracting the binary model from the light curve. We found that additional polynomial fits were needed to remove some non-periodic trends that remained. These trends were shorter in time than the de-trending windows used before. \jktebop\, allows addition of polynomial fits to account for variation of any parameter in the model. We added polynomials to account for the variation of the total light of the system.   We iteratively added polynomials to the model until we found the least residuals possible. We found that a good model only required two or three polynomials, without over-fitting any periodic trends.  After optimisation of all the above parameters, we optimised the limb-darkening coefficients ($u_1$, $u_2$) and also the third light fraction\footnote{we define third light ($l_3$) as per \cite{phoebeC2020}} (lf$_3$). The errors were calculated using the Monte Carlo (MC) algorithm available in \jktebop{} for 10000 iterations.  The final residuals give us the pulsation light curve, which is shown in the bottom panel of Fig.\ref{fig:phasedjklc}. 

While \jktebop{} is excellent for modelling the binary contribution to the light curve, we found that $r_{\rm 1}+r_{\rm 2} = 0.31$, indicating that the stars are close enough to have deformations that are better modelled using Roche geometry. We only used the \textsc{jktebop} model to extract the pulsation light curve as it was faster and easier to create an optimised binary model for all ten \textit{TESS} sectors, compared to other binary modelling codes.

\subsubsection{\phoebe}
To get a more accurate estimate of stellar parameters, we used the binary modelling code \phoebe{}\footnote{\url{https://phoebe-project.org/}} \citep{phoebe2016,phoebe2018,phoebeJ2020,phoebeC2020}. The presence of high-amplitude (orbit-synchronised) pulsations  may affect the measurement of a few parameters that are sensitive to the portions near the ingress and egress,  for example, limb-darkening coefficients, and radius ratio. Therefore, such systems need joint modelling of the pulsations, the eclipse, and the Roche deformations. No such setup is currently available. Therefore, we removed the two pulsations with the highest amplitude using pre-whitening. We then modelled this pulsation-subtracted light curve using version 2.4.17 of {\sc phoebe2}. 
We assigned the mean amplitude of the dominant pulsation as errors to the new pulsation-subtracted light curve. We do this to account for any loss of binary signature when removing the pulsations.
We used the light curve from the latest sectors: 83, 84, and 85. %
For computational efficiency, we randomly sampled the pulsation-subtracted light curve separately for eclipses and out-of-eclipse portions. We picked 1500 points from the eclipses and 500 from the out-of-eclipse portions.  We phased this resampled, and pulsation-subtracted light curve  over four orbital cycles and then proceeded to \phoebe\ optimisation using the Nelder-Mead algorithm \citep{NelderMead}.

\begin{figure}
    \includegraphics[width=\columnwidth]{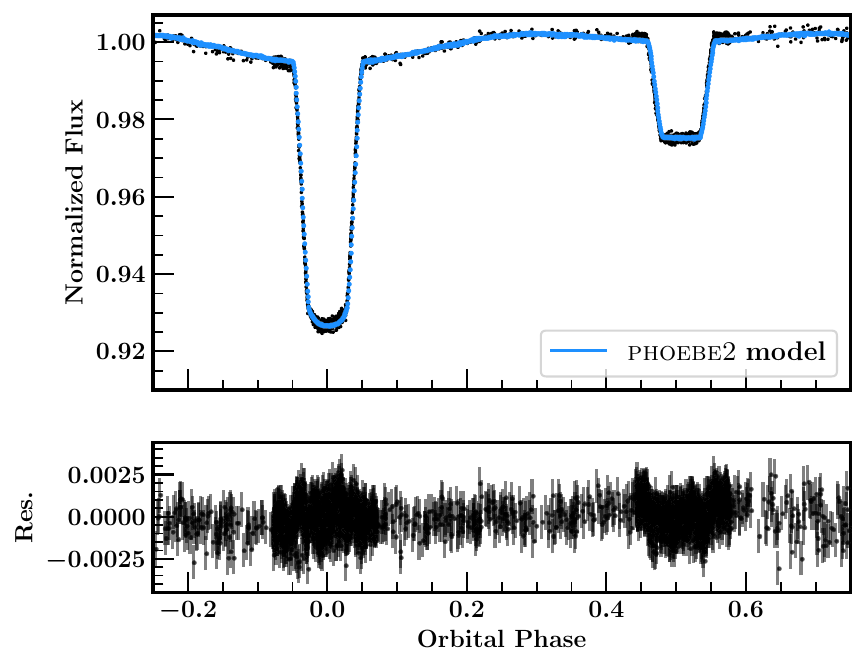}
    \caption{Phase-folded binary-LC (in black) with \phoebe{} model (in blue). The bottom panel shows the residuals of the fit along with the error bars of the data points, in black.}
    \label{fig:phasedphoebelc}
\end{figure}

We used initial values of the stellar and orbital parameters from the \jktebop{} solutions. The irradiation model was set to the Horvat scheme \citep{phoebe2016}. We set the gravity-darkening coefficients to the \phoebe\ suggested values of unity. Reflection coefficients for both stars were set to unity. We started with the limb darkening setup from \cite{HighMassPul_SouthBowm2022}. We fixed the primary star temperature to the initial estimates from spectral analysis. The $a_1\sin{i_1}$, and the mass ratio ($q_1$) were fixed to the estimates from the orbital solution from spectroscopy (see \autoref{tab:specorb}). After a loop of feedback between light curve modelling and spectroscopic analysis, we fixed the asynchronous rotation of the components (using the \texttt{syncpar} parameter) from the spectral analysis. We then sequentially optimised (in different runs) a set of parameters chosen from third light ($l_3$), $r_{\rm 1}+r_{\rm 2}$, $e_1$, $\omega_1$, $i_1$, $k$, $T_{0,1}$, $P_1$, passband luminosity of the primary ($L_\mathrm{pb,1}$), and the temperature ratio between the secondary star and the primary star ($T_\mathrm{ratio}$). The sequential optimisation was carried out until the root mean square (rms) of the residuals was less than 1~per~cent of the total flux of the system. 
The final optimised model had residuals of rms less than 0.25~per~cent of the total flux of the system (Fig.\ref{fig:phasedphoebelc}). The errors were sampled using the Markov Chain MC algorithm implemented in \phoebe. We used 60 walkers for 10 parameters (see \autoref{tab:phoebeparam}) and sampled it for 6000 steps with burn-in of 1000 steps. This sampling was carried out with a set of priors which included Gaussian distributions of effective temperature of the primary ($T_\mathrm{eff,1}$), $P_1$, $q_1$, $a\sin{i}$, and uniform distributions of the \texttt{syncpar} parameter for both the primary and secondary companion. The final \phoebe\, model is shown in Fig.\ref{fig:phasedphoebelc} and the MCMC corner plot can be found in Appendix \ref{appendix:mcmc}.

The final model gave us a $r_{\rm 1}+r_{\rm 2}$ =  0.31468$_{-0.00039}^{+0.00047}$ and a $k$ 
= 0.2583$_{-0.0017}^{+0.0015}$. 
We found a third light of 1.8~per~cent of the total light. The primary and secondary contribute around 96.3~per~cent and 1.9~per~cent of the total light,  respectively, with the secondary to primary light ratio being 0.0196. The final set of optimised parameters and their MCMC errors are given in \autoref{tab:phoebeparam}.

To check for the consistency of the solutions, we used the final MCMC values as the starting values for a Nelder-Mead optimisation for the sectors 16, 24, 56, and 76. These sectors we selected as they had the most data after removal of outliers and, together with the MCMC estimate, covered the most time. We subtracted the two dominant pulsations but did not resample these light curves. The light curves were optimised using Nelder-Mead for 150 iterations and involved the $P_1$, $T_{0,1}$, $r_{\rm 1}+r_{\rm 2}$, $i_1$, $e_1$, $\omega_1$, $k$, lf$_{3}$, and $T_\mathrm{ratio}$.  The final estimates from the Nelder-Mead optimisation were assigned the errors from the MCMC sampling. We noticed deviation in values of the Nelder-Mead optimised parameters but no more than three per cent of their corresponding MCMC estimate, except lf$_{3}$ where the MCMC estimate is less than half of the Nelder-Mead estimate (see Fig.\ref{fig:paramvar}). The difference could be due to a few number of iterations in the optimisation routine. Nonetheless, assuming that the MCMC errors were underestimated, we recalculated the parameter values and their errors by a weighted average of all the estimates: the Nelder-Mead estimates of sectors 16, 24, 56, 76, and the MCMC estimate. The MCMC estimate was assigned three times the weight compared to the Nelder-Mead estimates. The final values of the parameters are listed in \autoref{tab:phoebeparam}.\\

\begin{table}
    \centering
        \caption{\phoebe\, setup, priors, and the final parameters. The setup includes the \phoebe\, parameters corresponding to the limb darkening law, the two coefficients corresponding to the law, bolometric fraction of incident light that is reflected, coefficient for gravity darkening corrections, and the passband luminosity mode. The priors are included from the spectroscopic estimates where the \phoebe\, parameters for asynchronicity are calculated from the spectroscopic estimates of rotational velocities and the expected synchronised velocity. More details about the \phoebe\, parameters are available at:  \url{https://phoebe-project.org/docs/2.4/physics}.    }
\begin{tabular}{lcc}
\hline \hline
Parameter &  \multicolumn{2}{c}{Value}  \\
\hline
Fixed parameters   &      &                     \\
\hline
{\tt ld\_mode}  & \multicolumn{2}{c}{`quadratic'}\\
{\tt ld\_coeffs@primary}        & \multicolumn{2}{c}{ [0.094, 0.232] }  \\
{\tt ld\_coeffs@secondary}         &  \multicolumn{2}{c}{ [0.2,  0.238]}  \\
{\tt irrad\_frac\_refl\_bol}$^a$      & \multicolumn{2}{c}{ 1 }  \\
{\tt gravb\_bol}$^a$     &  \multicolumn{2}{c}{ 1 }  \\
{\tt pblum\_mode} &\multicolumn{2}{c}{ `component-coupled'} \\
\hline
Priors    &      &                     \\
\hline
$P_1$      [d]      &  \multicolumn{2}{c}{3.808567 $ \pm $ 0.00005  }   \\
$T_\mathrm{eff,1}$ [K]&   \multicolumn{2}{c}{ 24220 $ \pm $ 200  } \\
$q_1$           &    \multicolumn{2}{c}{0.1550 $ \pm $ 0.0012 }  \\
$a_1\,\sin i_1$ [R$_{\sun}] $  &  \multicolumn{2}{c}{23.555 $ \pm $ 0.05 }    \\
\texttt{syncpar@primary} &\multicolumn{2}{c}{ 1.81 $ \pm $ 0.2}\\
\texttt{syncpar@secondary}&  \multicolumn{2}{c}{ 3.15 $ \pm $ 0.3 }\\
\hline
\multicolumn{3}{l}{Fitted and sampled parameters}         \\
\hline
& MCMC estimate& Final value \\
\hline
  \vspace{0.1cm} 
  $T_0$ [BJD - 2457000]& \multicolumn{1}{c}{$3610.34690_{-0.00083}^{+0.00074}$} & \multicolumn{1}{c}{$3610.344 \pm 0.005$} \\
    \vspace{0.1cm}
  \vspace{0.1cm}
  $e_1$ & \multicolumn{1}{c}{$0.01981_{-0.00097}^{+0.00107}$} & 0.0190 $ \pm $ 0.0007 \\
      \vspace{0.1cm}
  $i_1$ [deg] & \multicolumn{1}{c}{$85.296_{-0.095}^{+0.091}$} & 85.9 $\pm$ 0.5  \\
      \vspace{0.1cm}
  $\omega_1$ [deg]& \multicolumn{1}{c}{$297.4_{-1.4}^{+1.5}$} & 298.3 $\pm$ 1.0 \\
  \vspace{0.1cm} 
$r_1+r_2$ &  $0.31468_{-0.00039}^{+0.00047}$ & 0.3131 $\pm$ 0.0015\\
\vspace{0.1cm} 
$k=r_2/r_1$ & $0.2583_{-0.0015}^{+0.0017}$ & 0.2610 $\pm$ 0.0026\\
\vspace{0.1cm} 
$L_\mathrm{pb,1}$ [W]& $12.27_{-0.15}^{+0.13}$& $12.27_{-0.15}^{+0.13}$\\
\vspace{0.1cm} 
$l_\mathrm{3}$ [${\rm Wm^{-2}}$]& $0.018_{-0.010}^{+0.014}$ & -\\
\vspace{0.1cm} 
lf$_{3}$ &  $0.018 \pm 0.012$$^b$& 0.034 $\pm$ 0.014 \\
$T_\mathrm{ratio}$ & $0.4630_{-0.0011}^{+0.0011}$ & 0.4632 $\pm$ 0.0016\\
\hline
\end{tabular}
\flushleft
$^a$ For all components \\  $^b$ Propagated from $l_3$ and  $L_\mathrm{pb}$
    \label{tab:phoebeparam}
\end{table}

\subsection{Spectral disentangling and analysis}

\subsubsection{Spectroscopic orbit}

The previous light curve \citep{HighMassPul_SouthBowm2022} and spectroscopic \citep{Tkachenko_2024} analyses made it clear that the secondary component in \vcep\ is a faint star, contributing barely 2~per~cent to the total light of the system. This was confirmed in our light curve analysis (c.f. \autoref{sec:lightcurvemod}). 

In this work, we employed the method of spectral disentangling (\spd; \citealt{Simon_Sturm_1994, Hadrava_1995}). \spd\ has some advantages for extracting spectra and radial velocities (RVs) in systems with two stars of different spectral types. First, there is no need for RV measurements as an input, hence any mismatches with template spectra are avoided. Second, the extracted spectra of the components do not assume any specific model template. This facilitates a detailed atmospheric analysis of the components. \spd\ is also a powerful method for the detection of faint component(s) in high contrast systems. Component(s) contributing only a few per~cent to the total light of the system have been successfully detected using \spd\ \citep{Mayer_2013, Torres_2014, Kolbas_2015, Themessl_2018, Johnston_2023}. This method has also been successful in cases where the spectral lines are almost washed out due to high rotational velocity and their depth is less than 1~per~cent of the normalised continuum \citep{Pavlovski_2022}. 

\begin{figure}
    \includegraphics[width=\columnwidth]{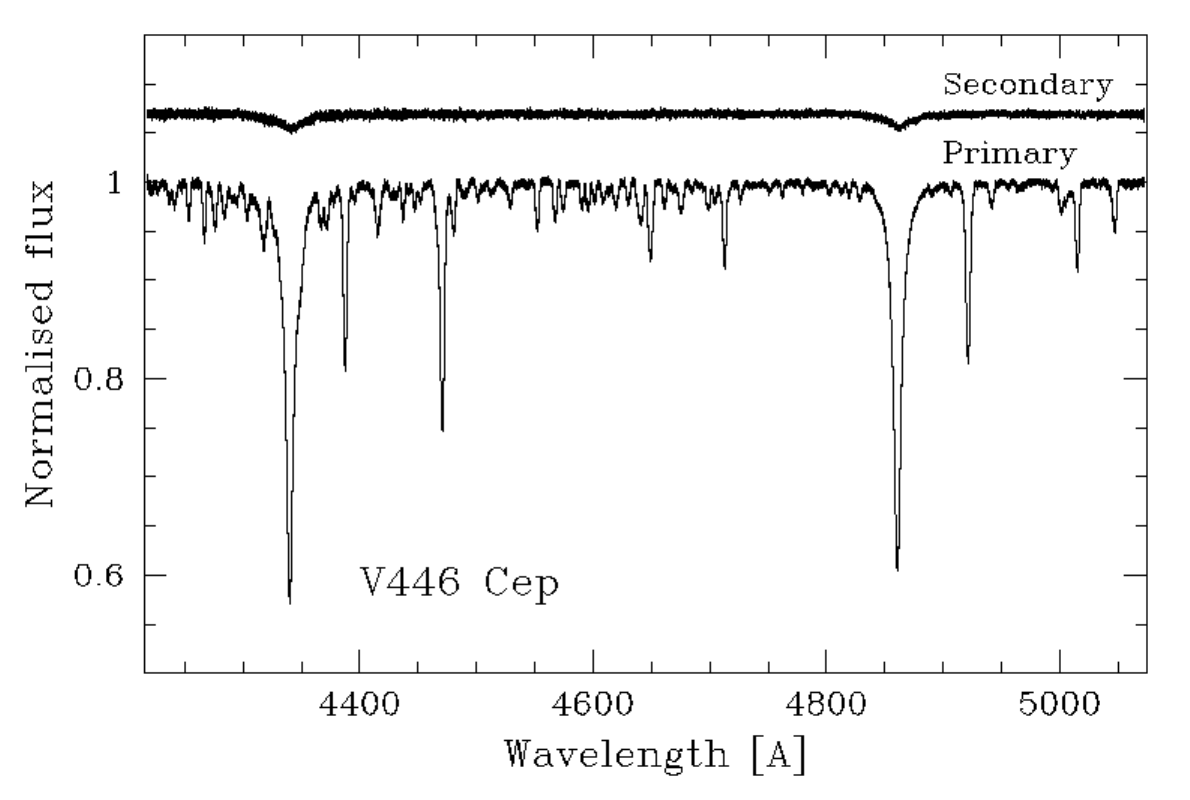}
    \caption{The spectra of the components in \vcep\ as reconstructed by \spd. Disentangled spectra are still in a common continuum of a binary system; hence, the highly diluted secondary component is obvious. }
    \label{fig:disent_spec}
\end{figure}

\begin{figure}
    \includegraphics[width=\columnwidth]{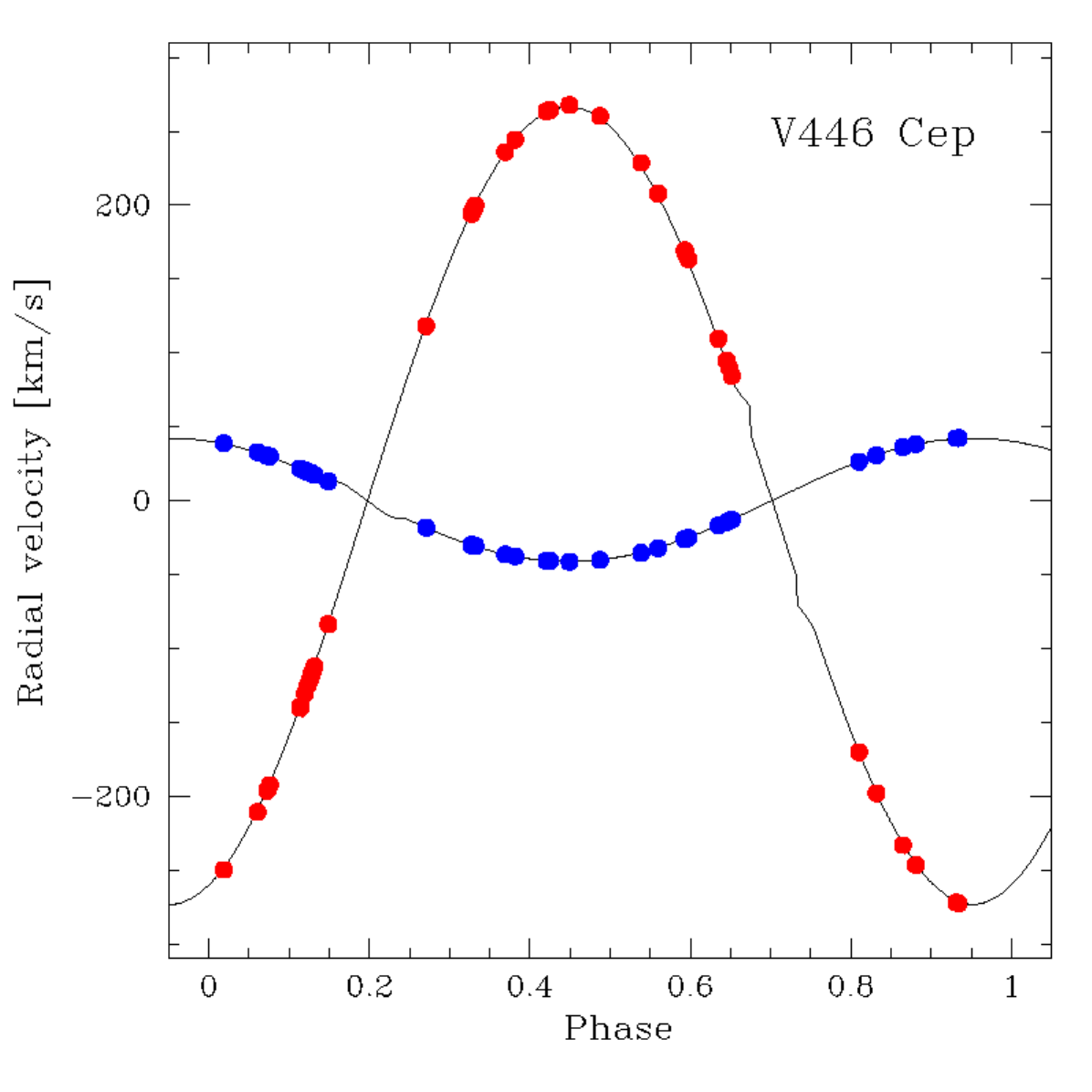}
    \caption{The solution for the spectroscopic orbit from \spd\ is represented in solid lines. The phase distribution of the observed spectra is illustrated with blue (primary) and red (secondary) circles. In \spd\ the orbital parameters are directly optimised by-passing the RVs for individual exposures - therefore symbols are only for illustrative purposes. The Rossiter-McLaughlin effect in the course of the primary and secondary eclipses is seen in the computed solution. }
    \label{fig:spec_orbit}
\end{figure}

One of the fundamental requirements of \spd\ is that there should be no intrinsic line profile variations (LPVs) in the spectra of the components, except for the dilution effect in the course of the orbital cycle. Such a variation can be due to the Rossiter-McLaughlin effect \citep{Kopal_1959}, which makes the spectra during the ingress and egress phases of an eclipse unreliable. In our sample of HERMES spectra, 33 were obtained during an eclipse. These spectra were not used in \spd, making the sample substantially smaller. Another source of LPVs is pulsations. Since \vcep\ contains a \bcep\, pulsating component, this is relevant in our analysis. It was found in practice that small line profile distortions due to non-radial modes would not seriously affect the RV curves \citep{Tkachenko_2012, Tkachenko_2014, Lehmann_2013, Themessl_2018, Johnston_2023} but if also radial pulsations are involved, another strategy in \spd\ is needed \citep{Tkachenko_2016, Pilecki_2021}.

For \spd\, we used the code {\sc FDBinary} \citep{Ilijic_2004} which is based on disentangling in Fourier space  \citep{Hadrava_1995} but implements a discrete Fourier transform (DFT). This enables more reliable handling of the input data. The methodology of \spd\ and subsequent atmospheric analysis adopted in the present work is extensively described in \citet{Pavlovski_2018,Pavlovski_2023}. 

After some trials with short spectral segments with a width $\sim 100$ {\AA}, which did not return any stable solutions, a longer segment spanning $\lambda =$ 4180 - 5150 {\AA} was adopted for the determination of the spectroscopic orbit with \spd. This segment includes two Balmer lines, H$\gamma$ and H$\beta$, some \ion{He}{i} lines, and numerous metal lines that are typical of early B-type stellar spectra. The projected rotational velocity of the primary component is high. As a consequence, the overlapping of spectral lines and line blends is present in the observed spectra. Usually, we avoid intrinsically broad Balmer lines in the determination of the orbital parameters: (i) the width of Balmer lines in B-type stars is larger than the RV changes for a given component, and (ii) broad Balmer lines in hot stars are usually extended to multiple \'{e}chelle orders. Proper normalisation of the observed spectra is challenging and could be a source of systematic errors in the determination of the orbital parameters. But in our case, we needed to include 
Balmer lines, along with the metal lines, for the faint companion. The light curve solution suggested that the secondary component is either a late B-type or an early A-type star with the effective temperature ($T_{\rm eff}$) around $11\,000$~K. Thus, its Balmer lines are almost at maximal strength. This gave us a much better prospect for extracting the secondary spectra compared to using only the the metal lines which are rather weak and not numerous, except for \ion{Mg}{ii} line at 4481~{\AA}.  The inclusion of Balmer lines for the secondary was proven to be a good decision since the spectrum of the secondary component was revealed to have Balmer line depths of only 1.5~per~cent of normalised flux, whilst depths of metal lines were found to be less than 0.3~per~cent and are affected by the noise (Fig.\ref{fig:disent_spec}). The parameters of the spectroscopic orbit are given in \autoref{tab:specorb}. The uncertainties of the orbital parameters were calculated with the bootstrap technique implemented by \citet{Pavlovski_2018}.

\begin{table}
\centering
\caption{Optimal parameters of the spectroscopic orbit determined by \spd. The period $P$ was fixed with the value found from the initial light curve solution. All other parameters of the orbit, time of periastron passage ($T_{\rm per}$), $e_1$, $\omega_1$, and the RV semi-amplitudes for both components, $K_1$ and $K_2$ were left free. $q_1$, $M_1\,\sin^3 i$ and $M_2\,\sin^3 i$ and the semimajor axes $a\,\sin i$, calculated from the optimised parameters, are given. }
\label{tab:specorb}
\begin{tabular}{ccc}
\hline \hline
Parameter & Unit & Value  \\
\hline
Determined    &      &                     \\
\hline
$P_1$           & [d]  & 3.808574 (fixed)    \\
$T_{\rm per}$ & [d]  & 55701.01$\pm$0.11   \\
$e_1$           &  -   & 0.0133$\pm$0.0021   \\
$\omega_1$      & [deg]& 74.0$\pm$9.8        \\
$K_1$         & [km\,s$^{-1}$] & 42.02$\pm$0.32  \\
$K_2$         & [km\,s$^{-1}$] & 271.05$\pm$0.49 \\
\hline
Calculated    &      &                         \\
\hline
$q_1$           & -    & 0.1550$\pm$0.0012       \\
$M_1\,\sin^3 i_1$ & M$_{\sun}$  & 10.603$\pm$0.057  \\
$M_2\,\sin^3 i_1$ & M$_{\sun}$  & 1.644$\pm$0.017   \\
$a_1\,\sin i_1$   & M$_{\sun} $  & 23.555$\pm$0.044     \\
\hline
\end{tabular}
\end{table}

The only spectroscopic orbits for \vcep\ in the literature are from \citet{Tkachenko_2024}. The agreement between 
our study and \citet{Tkachenko_2024}
is satisfactory except that the RV semi-amplitude of the secondary component ($K_2$) differs considerably. It is hard to trace the source of this inconsistency because the same observed spectra and the same disentangling code were used in both studies. However, we only used the 40 out-of-eclipse spectra. 
 This shows that LPV due to the Rossiter-McLaughlin effect (possibly coupled with the pulsations) could be severe enough to affect the orbital solution obtained using \spd. Further, in \citet{Tkachenko_2024}, the spectral segments avoided prominent Balmer lines because the normalisation routine, in the dedicated {\sc hermes} pipeline (version 7.0), was not ready to deal with the broad lines reliably. On the contrary, the detailed analysis of the system in this paper allowed us to use strong Balmer lines, along with the weak metal lines, for the \spd\, orbital solution.

The selection of the spectral segment also contributes to the discrepancy, as the line profiles vary substantially for different wavelength ranges (see Appendix \ref{appendix:BF} for examples).

While the value of $e_1$ we determined in \spd\ ($0.0133\pm0.0021$) is within the 3-$\sigma$ uncertainty of the value found in the light curve analysis ($0.0190\pm 0.0007$),  the discrepancy in $\omega_1$ between these two complementary solutions is large and hard to explain. In fact, $\omega_1$ is the most uncertain quantity in the spectroscopic solution. Apsidal motion, in spite of a very small eccentricity, might be a source of strong disagreement because of different epochs of the observations. Moreover, a 10-year gap between the observing seasons in which the spectra were obtained might be another reason for a large error in spectroscopic determination of $\omega_1$. 

In context of possible apsidal motion in \vcep\ we noticed that $\zeta$\,Phe, an intermediate mass binary system has very similar eccentricity, $e=0.0116\pm0.0024$ \citep{Southworth_2020} and one of the fastest apsidal motion with $\dot{\omega}= 6.16\pm 0.20~\mbox{deg}\,\mbox{yr}^{-1}$, corresponding to the period $U= 58.4 \pm 1.8$~yr \citep{Zasche_Wolf_2007}. The fastest apsidal motion was found in GL Car with $\dot{\omega}= 14.29 \pm 0.03~\mbox{deg}\,\mbox{yr}^{-1}$, and  $U= 25.20 \pm 0.02$~yr with a higher eccentricity \citep{Wolf_2008}. An apsidal motion of $\dot{\omega} \sim 22~\mbox{deg}\,\mbox{yr}^{-1}$ ($15~\mbox{deg}\,\mbox{yr}^{-1}$ for decreasing $\omega_1$) would explain the $\omega_1$ discrepancy in \vcep, but such a fast apsidal motion was not detected in the eclipse timing variations (ETVs; see \autoref{sec:etv}). But the ETVs point at variations of the time of periastron which could be a reason for the discrepancy. We did a check for a \spd\, solution with an additional companion but could not get a solution. Sometimes there is an ambiguity between solutions differing by $\pm \pi$. Unfortunately we can not resolve the discrepancy for our results with certainty.

\subsubsection{The primary's spectrum in the total eclipse}
\label{sec:eclipsespe}

\begin{figure}
    \includegraphics[width=\columnwidth]{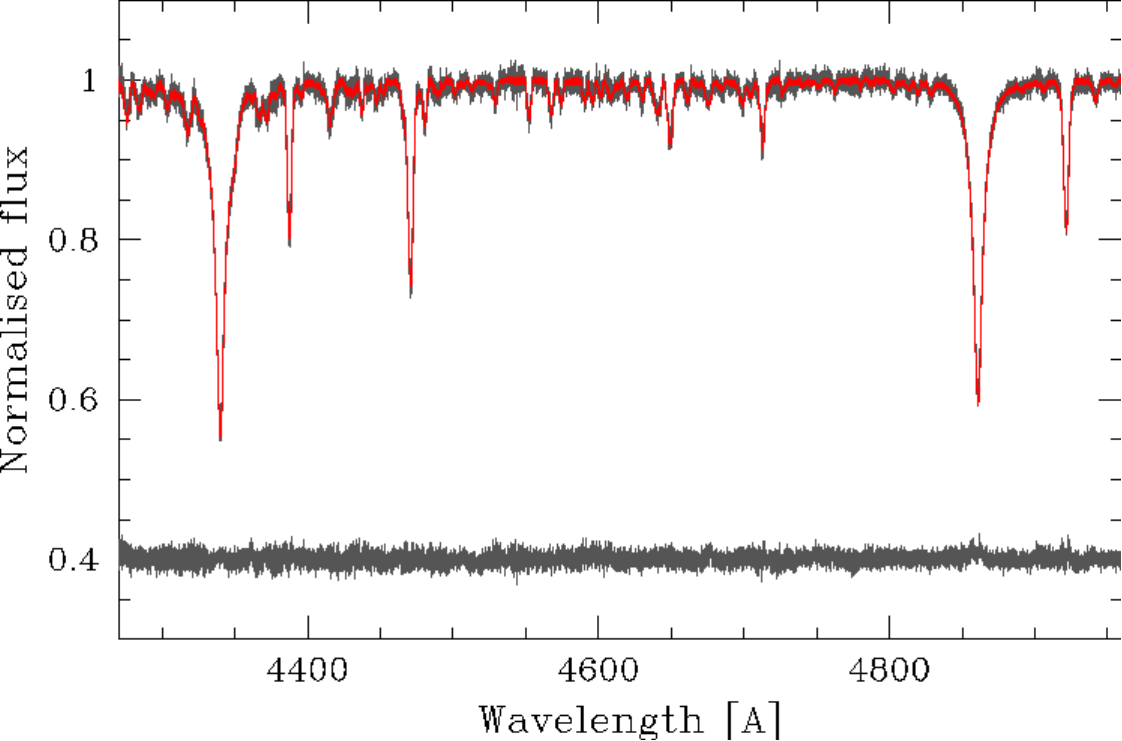}
    \caption{Disentangled spectrum of the primary component (in red) superimposed on the primary's spectrum (in black) obtained in the total eclipse. The residuals are also shown, and shifted by 0.4 in the units of the normalised continuum for convenience.}
    \label{fig:disent_total_spe}
\end{figure}

The secondary eclipse in the light curve of \vcep\ is total with the primary (i.e.\ hotter and brighter) component eclipsing its companion. This allows direct comparison of the primary's spectrum with its disentangled spectrum. Most of the eclipse spectra were obtained in the phases of ingress or egress of the eclipses, but two  of the spectra are very close to the midpoint of the total eclipse. These two spectra resemble each other, so we decided to co-add them to enhance the signal-to-noise ratio (S/N).

The disentangled spectrum of the primary component is diluted, and for direct comparison to the eclipse spectrum, the light ratio should be known. We turned this around and determined the light ratio by varying it until these two spectra match, minimising the sum of the squared residuals. This gave us $l_{\rm s}/l_{\rm p} = 0.022\pm0.004$, which is in excellent agreement with the light curve analysis. The slight difference could be due to the wavelength dependence of the light ratio since the spectra belong to the blue spectral region whilst \textit{TESS} is sensitive to redder wavelengths. The comparison of the scaled disentangled spectra and the eclipse spectra is shown in Fig.\ref{fig:disent_total_spe}. No conspicuous trends in the residuals are noticeable. 

Since \spd\ co-adds all observed spectra, the disentangled spectra of the components gain in S/N. 
We measured S/N in a short spectral segment free of the spectral lines from 4570 {\AA} to 4580 {\AA}, the region of the \ion{Si}{iii} triplet. This yielded S/N $= 670$ for the primary's disentangled spectra compared to the S/N of 155 for the eclipse spectra.

\subsubsection{Atmospheric parameters}

\begin{figure}
    \includegraphics[width=\columnwidth]{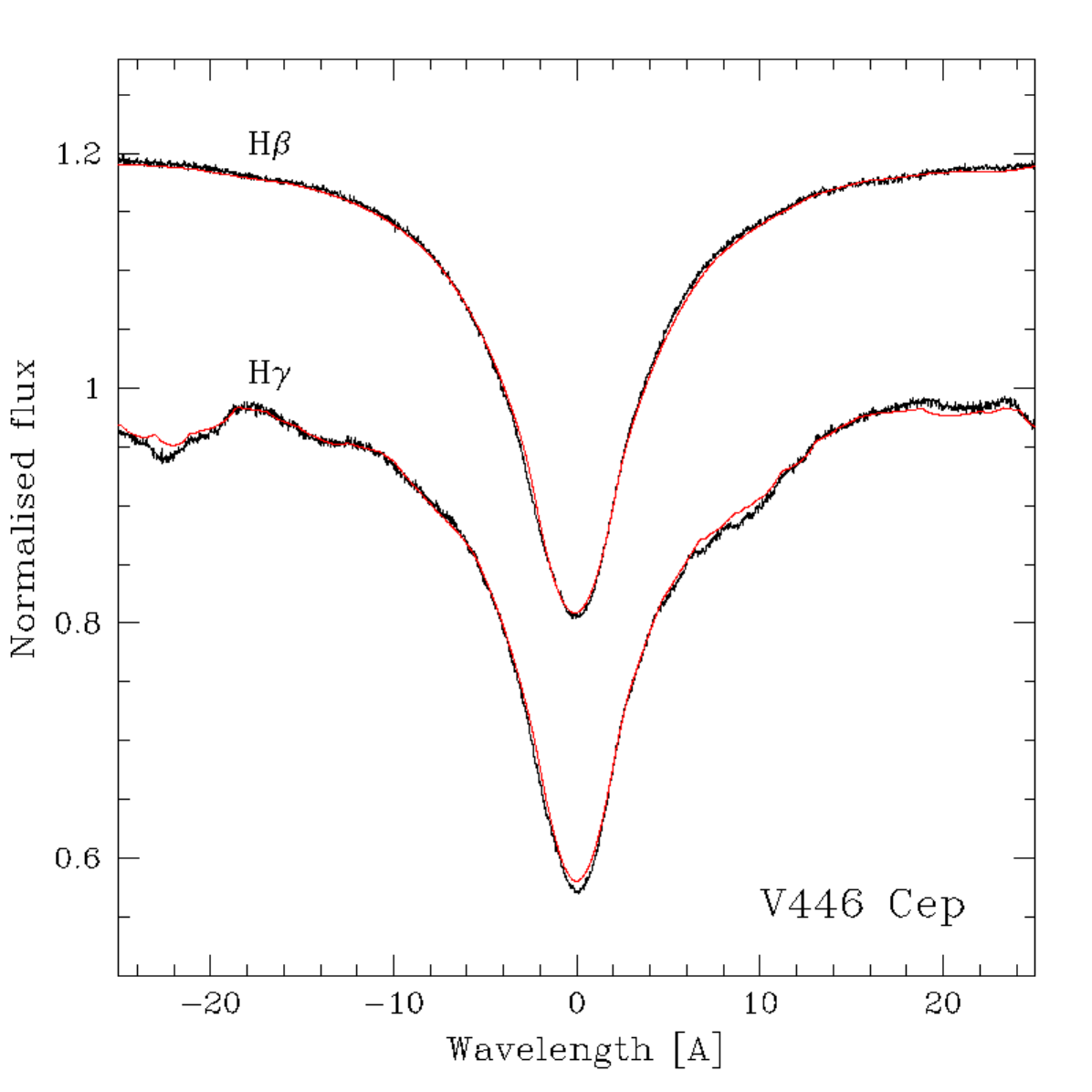}
    \caption{Optimal fits (in red) of the H$\gamma$ and H$\beta$ in the disentangled spectrum of the primary component of V446 Cep (in black). In fitting the disentangled spectrum, metal lines superimposed on the wings of the Balmer lines were masked. The $\log g$ and the light ratio were fixed from the solution of the light curve analysis.}
    \label{fig:balmer.fits}
\end{figure}

\begin{table}
\centering
\caption{Atmospheric parameters determined for the components of the binary system \vcep. }
\label{tab:atmosfit}
\begin{tabular}{lccc}
\hline \hline
Parameter      & Unit & Primary  & Secondary \\    
\hline
Constrained    &      &    &                 \\
\hline
$T_{\rm eff}$  & [K]  &  23830$\pm$260  &  9080$\pm$390                  \\
$\log g$ & [cgs]  & 3.91 (fix) & 4.30 (fix)  \\
lf          & & 0.978$\pm$0.006  &    0.022$\pm$0.003       \\        
$v\sin i_1$      & [km\,s$^{-1}$] & 143.2$\pm$3.0  &    64.1$\pm$6.3    \\
\hline

Eclipse        &      &       &              \\
\hline
$T_{\rm eff}$  & [K]  &  24220$\pm$180     &              \\
$\log g$ & [cgs]  & 3.91 (fix) &             \\
lf             &  -   &  1.0 (fix)      &             \\         
$v\sin i_1$      & [km\,s$^{-1}$] & 142.5$\pm$2.9    &      \\
\hline
\end{tabular}
\end{table}

The light curve analysis and the spectroscopic orbit gave us multiple constraints for our spectroscopic analysis. For example, hydrogen lines are excellent indicators of the $T_{\rm eff}$ with the drawback that for stars with $T_{\rm eff} > 8000$~K, a strong degeneracy appears with the surface gravity ($\log g$). However, the $\log g$ of the components of DLEBs can be measured with high precision.
Thus, we lifted the degeneracy between the $T_{\rm eff}$ and $\log g$ by fixing $\log g$  in our estimation. 

The spectrum of the primary component was also characterised by including its strong \ion{He}{i} lines, specifically for constraining $T_{\rm eff}$. They were also useful for the determination of the projected rotational velocity ($v\,\sin\,i$), which cannot be determined by only using the Balmer lines. We also found one weak line of \ion{He}{ii} at 4686~{\AA} which gave an upper limit to the primary's $T_{\rm eff}$. 
In the spectrum of the secondary component, we see the strong H$\gamma$ and H$\beta$ Balmer lines along with some faint metal lines. The renormalised spectrum of the secondary component yielded S/N $\simeq 14$. Further, the large difference in the $T_{\rm eff}$  of the components meant that we had to use two grids of the theoretical spectra, one with local thermodynamic equilibrium (LTE) approximation for the secondary component, and the other with non-LTE (NLTE) assumption for the primary component.

The code {\sc starfit} \citep{Kolbas_2015} was used in the determination of the optimal atmospheric parameters for the components of \vcep. In {\sc starfit}, the disentangled and renormalised spectra are compared to pre-calculated theoretical spectra. The optimisation is performed using a genetic algorithm with the subroutine {\sc amoeba} \citep{Charbonneau_1995}. The uncertainties are calculated with MCMC sampling \citep{Foreman_Mackey_2013}. Details on the grids with theoretical LTE spectra are given in \citet{Kolbas_2015}, and for NLTE spectra in \citet{Pavlovski_2018, Pavlovski_2023}. Both grids are calculated for solar metallicity, [M/H] $=$ 0, which eventually limits the scope of our results.

In our calculations, the metal lines in the primary's disentangled spectrum were masked, along with the wings of H$\gamma$ and H$\beta$ lines. For the secondary's disentangled spectrum, we fit only spectral segments with a width of about 80 {\AA} and centred on the H$\gamma$ and H$\beta$ lines. Spectral fitting was performed in a constrained mode with fixed values of $\log g$ for both components. The atmospheric parameters obtained and the light dilution factors\footnote{with the condition $l_1 + l_2 = 1$} are given in \autoref{tab:atmosfit}. The quality of the best fit is shown in Fig.\ref{fig:balmer.fits} for the primary's H$\gamma$ and H$\beta$ lines, respectively. We also fit the eclipse spectra to check for any difference arising from our method of renormalising disentangled spectra. We find that, even with much lower S/N than in the primary's disentangled spectrum, the eclipse spectra give similar parameters, which agrees within the 1-$\sigma$ uncertainties (\autoref{tab:atmosfit}). 

Finally, we note that the $v\,\sin\,i$ for both components, $v_1\sin i_1 = 143.2\pm3.0$~km\,s$^{-1}$ and $v_2\sin i_1 = 64.1\pm6.3$~km\,s$^{-1}$, are substantially larger than the synchronised rotational velocities inferred from the components' radii, which are $v_1\sin i_{\rm 1,syn} = 76.8\pm0.3$~km\,s$^{-1}$ and $v_2\sin i_{\rm 1,syn} = 19.9\pm0.1$~km\,s$^{-1}$, respectively. 

\subsubsection{Abundances}

The high S/N of the disentangled spectrum of the primary component allowed us to determine the individual abundances for several atomic species in its photosphere. Its early-B-type characteristics mean that the lines of \ion{C}, \ion{N}, \ion{O}, \ion{Mg}\, and \ion{Al}\, are mostly present in a single ionisation stage. 

A hybrid NLTE model approach was adopted where modelling combines hydrostatic, plane-parallel, and line-blanketed model atmospheres in LTE with line formation calculated in NLTE \citep{Nieva_2007, Nieva_Przybilla_2012}. The {\sc Atlas9} code \citep{Kurucz_1979,Castelli_Kurucz2003} was used in the calculations of model atmospheres. Then emergent fluxes and line profiles were calculated with the codes {\sc Detail} and {\sc Surface} \citep{Giddings_1980,Butler_1984}. In {\sc Detail}, coupled radiative transfer and statistical equilibrium equations were solved, while {\sc Surface} was used for the calculations of NLTE synthetic spectra. The model atoms used were the result of a  dedicated critical evaluation by K.\ Butler (Munich), N.\ Przybilla (Innsbruck), and their associates over decades, with the most recent list of references for different ions given by \citet{Aschenbrenner_2023}. 

In the calculation of the model atmosphere for the primary component, we set $T_{\rm eff} = 23830$~K and $\log g = 3.91$. The grid of synthetic spectra was calculated for different abundances in steps of 0.05 dex, and microturbulent velocity $\xi$ in steps of 1~km\,s$^{-1}$. Finally, the calculated spectra were  broadened by a rotational profile  with $v\sin i = 143$~km\,s$^{-1}$. The least scatter in oxygen abundance was found for $\xi = 4\pm1$~km\,s$^{-1}$, which was then used in the abundance determinations for other species. The results of the abundance determination is given in \autoref{tab:abund}. The abundance of helium was not determined since we used the helium lines in the determination of the $T_{\rm eff}$. 
In the model atmosphere for the primary, the He abundance was set to the standard value $\log \epsilon({\rm He}) = 0.089$ by number of atoms. We found that the abundance patterns in \vcep\ are consistent with those of OB binaries \citep{Pavlovski_2018,Pavlovski_2023} and even single B-type stars \citep{Nieva_Przybilla_2012}. We could not extract abundances for the secondary as we had limited lines.

\begin{table}
\centering
\caption{Elemental abundances for the primary component of \vcep. The mean values found for the sample of 22 stars in OB binaries from \citet{Pavlovski_2018, Pavlovski_2023} and for `present-day cosmic abundances' of single sharp-lined early B-type stars determined by \citet{Nieva_Przybilla_2012} are given for comparison. The number in square brackets represents the number of lines considered for the analysis.}
\label{tab:abund}
\begin{tabular}{lcccc}
\hline \hline
Element & \vcep\ A & No. of lines & OB binaries & B-type stars  \\
\hline
$\log  \epsilon{\rm (C)}$    &  8.16$\pm$0.11 & 7    & 8.25$\pm$0.07   & 8.33$\pm$0.04   \\
$\log  \epsilon{\rm (N)} $   & 7.68$\pm$0.10 & 21    &  7.69$\pm$0.06  & 7.79$\pm$0.04   \\
$\log  \epsilon{\rm (O)} $   & 8.67$\pm$0.12 & 15     & 8.71$\pm$0.05   &  8.76$\pm$0.05  \\
$\log  \epsilon{\rm (Mg)} $   & 7.60$\pm$0.09 & 2     &   7.56$\pm$0.12 &  7.56$\pm$0.05  \\
$\log  \epsilon{\rm (Si)} $   &  7.33$\pm$0.16 & 3    & 7.45$\pm$0.09    & 7.50$\pm$0.05   \\
$\log \epsilon({\rm N/C})$ & -0.48$\pm$0.15 & - &-0.56$\pm$0.08 & -0.54$\pm$0.06 \\
$\log \epsilon({\rm N/O})$ & -0.99$\pm$0.13  & - &-1.02$\pm$0.07 & -0.97$\pm$0.06 \\
\hline
\end{tabular}
\end{table}


The final set of stellar, atmospheric, and orbital parameters that we adopt are given in \autoref{tab:paramtable}. We use the light curve estimates for $P_1$, $T_{0,1}$, $i_1$, $e_1$, and $\omega_1$, while we adopt the spectroscopic mass ratio. $i_1$, $k$, and $r_1+r_2$ from light curve solution were used with the spectroscopic solution to obtain the stellar parameters. We adopted the $T_{\rm eff}$ and $v\mathrm{sin}(i_1)$ from the eclipse spectra for the primary. The only possible direct measurement of the secondary's $T_{\rm eff}$ and $v\mathrm{sin}(i_1)$ was available from the disentangled spectra.

\begin{table}
 \centering
 \caption{Adopted parameters of the \vcep{} binary system. We adopt $e_1$, and $\omega_1$ from the light curve solution. $a_1$ was calculated from $a_1 \sin{i_1}$ obtained using \spd, and $i_1$ from light curve solution.   }
 \label{tab:paramtable}
 \small
\begin{tabular}{@{}lll}
\hline
\hline
\multicolumn{3}{c}{Orbital Parameters} \\
\hline
   & \multicolumn{2}{c}{A--B}    \\
  \hline
     \vspace{0.1cm} 
  $T_{0,1}$ [BJD - 2457000]& \multicolumn{2}{c}{3610.34690(83)} \\
    \vspace{0.1cm}
  $P_1$ [d] & \multicolumn{2}{c}{3.808567(12)}  \\
  \vspace{0.1cm}
  $a_1$ [R$_\odot$] & \multicolumn{2}{c}{23.62(5)}\ \\

      \vspace{0.1cm}
  $i_1$ [deg] & \multicolumn{2}{c}{85.9(5)}   \\
    \vspace{0.1cm}
  $e_1$ & \multicolumn{2}{c}{0.0190(7)}  \\
      \vspace{0.1cm}
  $\omega_1$ [deg]& \multicolumn{2}{c}{298(1)}   \\
      \vspace{0.1cm}
  $q_1$ & \multicolumn{2}{c}{0.1550(12)}  \\
 \hline
\multicolumn{3}{c}{Stellar and atmospheric parameters} \\
\hline
   & A & B  \\
   \hline
    \vspace{0.1cm}
 $M$ [M$_\odot$] & 10.68(6) &1.657(17) \\
 \vspace{0.1cm}
 $R$ [R$_\odot$] & 5.864(33)  &1.530(14)  \\
  \vspace{0.1cm}
 $T_\mathrm{eff}$ [K]&  24220(180) & 9080(390) \\
  \vspace{0.1cm}
 $\log(g)$ [dex] &3.946(5) &4.303(9)\\
 $v\mathrm{sin}(i_1)$ [km~s$^{-1}$] &  143(3)  &    64(6)  \\
 $u_1$ & 0.094 & 0.200 \\
 $u_2$ & 0.232 &0.237 \\
  lf (light curve) & 0.947(17) & 0.019(12) \\
   \vspace{0.1cm}
  lf (spectroscopy) & 0.978(6) & 0.022(3) \\
  Fractional radii &    0.2483(13)    &  0.0648(6)   \\
     \vspace{0.1cm}
  $T_\mathrm{ratio}$  light curve  &\multicolumn{2}{c}{0.4632(16)} \\

 $T_\mathrm{ratio}$ spectroscopy  &\multicolumn{2}{c}{0.375(19)} \\

 \hline
\multicolumn{3}{c}{System parameters} \\
  \hline
  $\mathrm{[M/H]}$  [dex]    &\multicolumn{2}{c}{0 (assumed)} \\
  \vspace{0.1cm}
Age [Myr] &\multicolumn{2}{c}{14.5(5)}\\
  \vspace{0.1cm}
lf$_{3}$  & \multicolumn{2}{c}{0.034(14)} \\

Distance [pc]   &\multicolumn{2}{c}{750.0$^{a}$} \\  
\hline
\end{tabular}

$^a$ From SED (see \autoref{sec:evo}).  

\end{table}

\section{Frequency Analysis}
\label{sec:pulsations}
\subsection{Frequency extraction}

We created the pulsation light curve by stitching the residuals from sectors 16-85 after \jktebop{} modelling. This resulted in a dataset of 128,537 points and a timebase of 1846~d. We used the discrete Fourier transform (DFT) and extracted the pulsation frequencies using \textsc{period04} \citep{Lenz_2005}. We looked in the frequency range between 0 and 30 d$^{-1}$. In an initial check, we did not find any frequencies beyond this range with sufficient S/N that were recurrent in all sectors. Hence, all our analyses are confined to this range.

\begin{figure}
    \centering
        \includegraphics[width=\columnwidth]{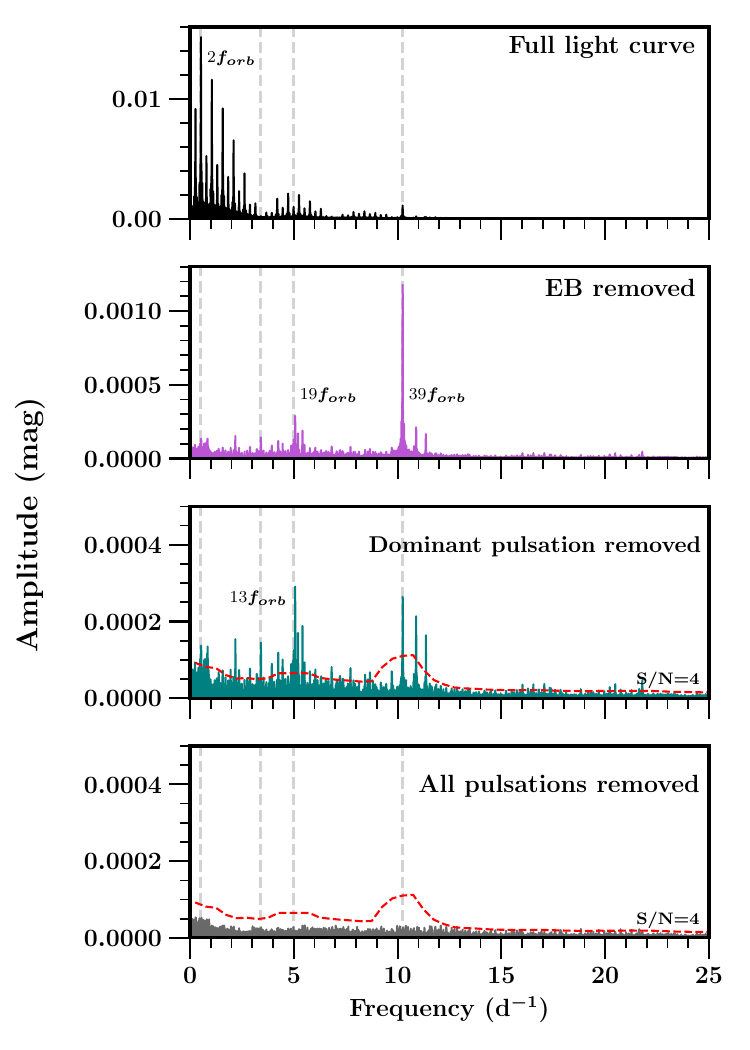}
    \caption{DFT of the light curves for different stages of frequency extraction. The top panel shows the DFT (in black) of the full \textit{TESS} light curve with eclipses. The panel second from the top (in purple) shows the DFT of the eclipse-subtracted light curve. The second from the bottom (in green) is the pre-whitened DFT after the removal of the dominant pulsation frequency. The bottom-most panel shows the pre-whitened DFT after the removal of all 142 pulsations. The grey lines denote different harmonics of the orbital frequency. The red dashed line shows the limit for an S/N of 4. }
    \label{fig:dfts}
\end{figure}

We extended the search beyond the suggested S/N limit for TESS data \citep{Baran_Koen21} to search for low-amplitude multiplets, if present.
We extracted all significant frequencies with iterative pre-whitening until we reached the S/N criterion of 4 (Fig.\ref{fig:dfts}).

There were long gaps in the total light curve, but four sets of consecutive sectors were available. To check for the consistency of the frequency solution, we divided the light curve into sets of consecutive sectors: 16-17, 56-57, 76-77, and 83-85. We compared the frequency spectrum for these sets and checked for recurrent frequencies and ruled out binary signatures and aliases. We accepted or rejected frequencies based on the methodology detailed in Appendix \ref{appendix:pulsefilter}. This gave us 21 recurrent frequencies. 

The final set of frequencies, calculated from the whole dataset, is listed in \autoref{tab:freq}. We checked if these frequencies are harmonics of the orbital frequency ($f_\mathrm{orb}$) or combinations of each other. We considered the combinations that were visible in all sets of consecutive sectors. Using the Rayleigh criterion of 0.012 d$^{-1}$ (for two {\it TESS} sectors), we found only three such possibilities as listed in \autoref{tab:freq}.

\begin{table*}
    \caption{Final set of extracted pulsation frequencies, which are listed in order of decreasing amplitude. We also list the closest orbital harmonic to each frequency in terms of N, where the harmonic frequency is N$\times f_\mathrm{orb}$.  To see which frequencies could be a harmonic, we list $(f-N\times f_\mathrm{orb})\times \Delta T $, where $\Delta T$ is the total length of the data set in days.   Possible combination frequencies are also listed in the table.  }
\begin{tabular}{cccccccc}
\hline
\hline
No. & $f (\mathrm{d}^{-1})$ & Amplitude (mmag) & Phase (rad) & S/N & N & $(f-N\times f_\mathrm{orb})\times \Delta T $& Remarks \\
\hline
$f_{1}$ & 10.2432419(5) & 1.20(1) & 0.201(2) & 46 & 39 & 5.6(6) & - \\
$f_{2}$ & 5.064302(2) & 0.30(1) & 0.618(5) & 20 & 19 & 139.3(3) & - \\
$f_{3}$ & 10.893687(2) & 0.23(1) & 0.069(7) & 10 & 41 & 236.9(6) & - \\
$f_{4}$ & 5.417558(3) & 0.18(1) & 0.435(8) & 12 & 20 & 306.7(3) & - \\
$f_{5}$ & 11.362614(3) & 0.17(1) & 0.400(9) & 11 & 43 & 133.1(7) & - \\
$f_{6}$ & 5.201641(3) & 0.16(1) & 0.779(9) & 11 & 19 & 392.9(3) & - \\
$f_{7}$ & 2.182762(3) & 0.15(1) & 0.82(1) & 12 & 8 & 151.7(1) & - \\
$f_{8}$ & 4.251804(4) & 0.12(1) & 0.93(1) & 8 & 16 & 93.6(2) & - \\
$f_{9}$ & 4.952164(6) & 0.09(1) & 0.41(2) & 6 & 18 & 417.0(3) & - \\
$f_{10}$ & 7.730014(6) & 0.09(1) & 0.69(2) & 9 & 29 & 213.2(4) & - \\
$f_{11}$ & 4.866921(7) & 0.08(1) & 0.76(2) & 5 & 18 & 259.7(3) & - \\
$f_{12}$ & 6.318200(8) & 0.07(1) & 0.48(2) & 6 & 24 & 30.5(4) & - \\
$f_{13}$ & 7.231708(8) & 0.06(1) & 0.86(3) & 6 & 27 & 262.7(4) & - \\
$f_{14}$ & 3.41200(1) & 0.05(1) & 0.52(4) & 5 & 12 & 482.1(2) & - \\
$f_{15}$ & 7.06233(1) & 0.05(1) & 0.05(3) & 5 & 26 & 434.8(4) & - \\
$f_{16}$ & 7.20783(1) & 0.05(1) & 0.31(3) & 5 & 27 & 218.6(4) & $ f_{10}$-2$f_\mathrm{orb}$ \\
$f_{17}$ & 2.75414(1) & 0.05(1) & 0.11(3) & 4 & 10 & 237.1(2) & - \\
$f_{18}$ & 8.57942(1) & 0.04(1) & 0.56(4) & 4 & 32 & 327.1(5) & - \\
$f_{19}$ & 21.78854(1) & 0.04(1) & 0.20(4) & 10 & 82 & 476(1) & $ 2f_{3}$ \\
$f_{20}$ & 20.48212(2) & 0.03(1) & 0.84(5) & 9 & 78 & 3(1) & $ 2f_{1}$ \\
$f_{21}$ & 17.39559(3) & 0.02(1) & 0.07(8) & 5 & 66 & 122(1) & - \\
\hline

\end{tabular}

    \label{tab:freq}
\end{table*}

 In an attempt to identify the pulsation modes $f_1$ and $f_2$, we tried to check for LPV using the HERMES spectra. The total dataset comprises of 72 spectra with a median temporal sampling of approximately $0.12\,\mathrm{d^{-1}}$, corresponding to a Nyquist frequency limit near $4.2\,\mathrm{d^{-1}}$. Compared with the most prominent pulsation frequencies found from \textit{TESS}, i.e., $f_1 = 10.24324\,\mathrm{d^{-1}}$ and $f_2 = 5.06430\,\mathrm{d^{-1}}$, this Nyquist limit imposes a physical constraint beyond which frequency information is aliased, complicating the interpretation of pulsation modes. The spectra were also taken in two bunches with a 10-year gap. 
   Despite these limitations, one may still gain insights by focusing on trends in amplitude and phase changes across stellar spectral lines as a function of velocity or wavelength, using the intensity period search \citep[IPS;][]{1997A&A...317..723T} method. We found that the LPV were affected by observational noise, especially for Si\,III 4552.62\,\AA\, and Si\,III 4567.872\,\AA\ lines, which are well-known diagnostics of pulsations in early-type stars due to their sensitivity to velocity fields induced by stellar oscillations \citep[e.g.,][]{2003A&A...401..281B,2005MNRAS.362..619B}.  We also tried to use the strong He\,I 4026\,\AA\ line, but the amplitude and phase variations were not conclusive.

\subsection{Tracing tidal effects}

The dominant pulsation frequency is a near-harmonic of the orbital frequency ($f_1 = 39 f_{\rm orb} +0.003~\mathrm{d}^{-1}$) and is visible in the phase-folded residuals of the \jktebop{} light curve fitting, shown in Fig.\ref{fig:phasedjklc}. 

\begin{figure}
    \centering
    \includegraphics[width=\columnwidth]{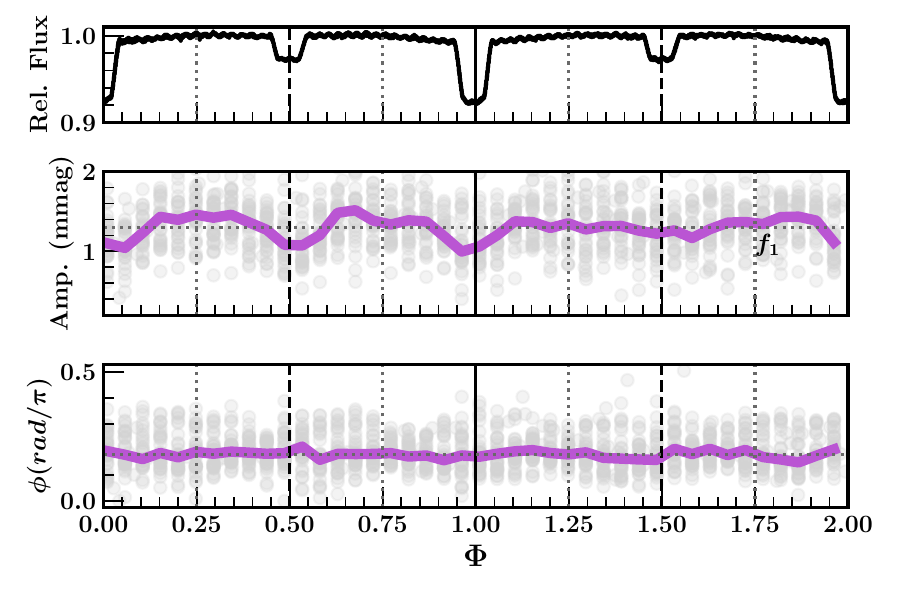}
        \includegraphics[width=\columnwidth]{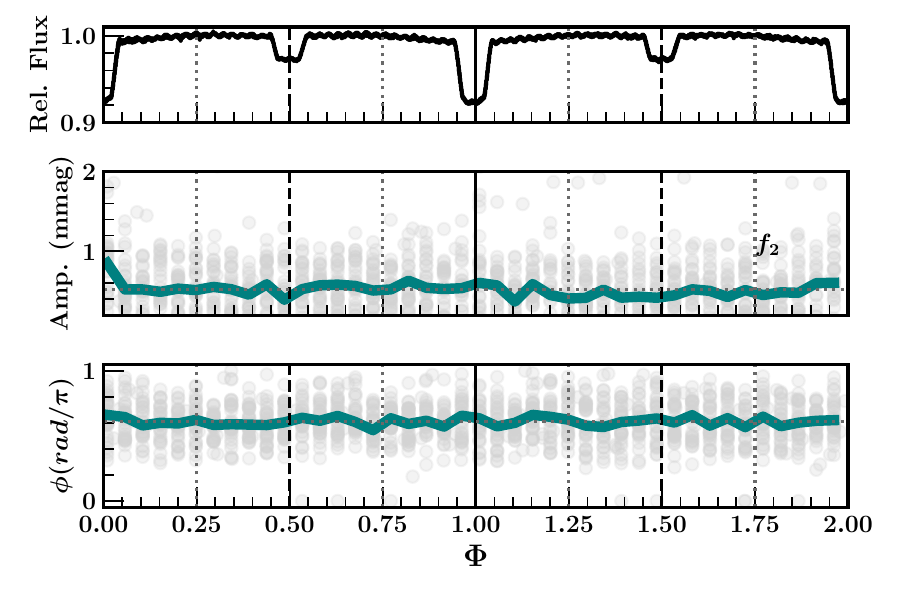}
    \caption{Pulsation runs for $f_1$ (top; purple), and $f_2$ (bottom; green). The dominant pulsation frequency, $f_1$, has slight amplitude variations at primary and secondary eclipses, but not substantial enough to confirm tidal perturbation. }
    \label{fig:f1run}
\end{figure}

This near-harmonic frequency could be a result of possible interaction between the stars and the pulsation. To check if this frequency is tidally perturbed, we need to observe variation in the mode's amplitude and phase over the orbital motion. We look for amplitude and phase variation of this frequency by following an approach similar to \cite{Steindl2021}. We divided the pulsation light curve into small segments that corresponded to 21 divisions of the orbital phase. We then fit the small segments separately with the function

\begin{equation}
    A(t)=A_0 \sin{(2\pi(f_i  t + \phi))}+c_0 ~ ,
\end{equation}
where we fix the frequency $f_i$, and fit for the amplitude $A_0$, and $\phi$ of each segment, and $c_0$ is the zero-point shift in the amplitude of the fit. We use the Levenberg–Marquardt algorithm to fit for the parameters with constraints on $\phi$ to vary from 0 to 1.
The $A_0$ and $\phi$ variations for $f_1$ and $f_2$, as a function of orbital phase ($\Phi$), are shown in Fig.\ref{fig:f1run}. We notice that $f_1$ has low amplitudes at both eclipses. But these amplitude variations are small in scale and therefore do not provide any strong evidence of tidal perturbation. The phase does not even vary slightly over the orbital phase. Meanwhile, for $f_2$, we do not see any variations in either amplitude or phase. 
 
\begin{figure}
        \vspace{-0.5cm}
    \includegraphics[width=\columnwidth]{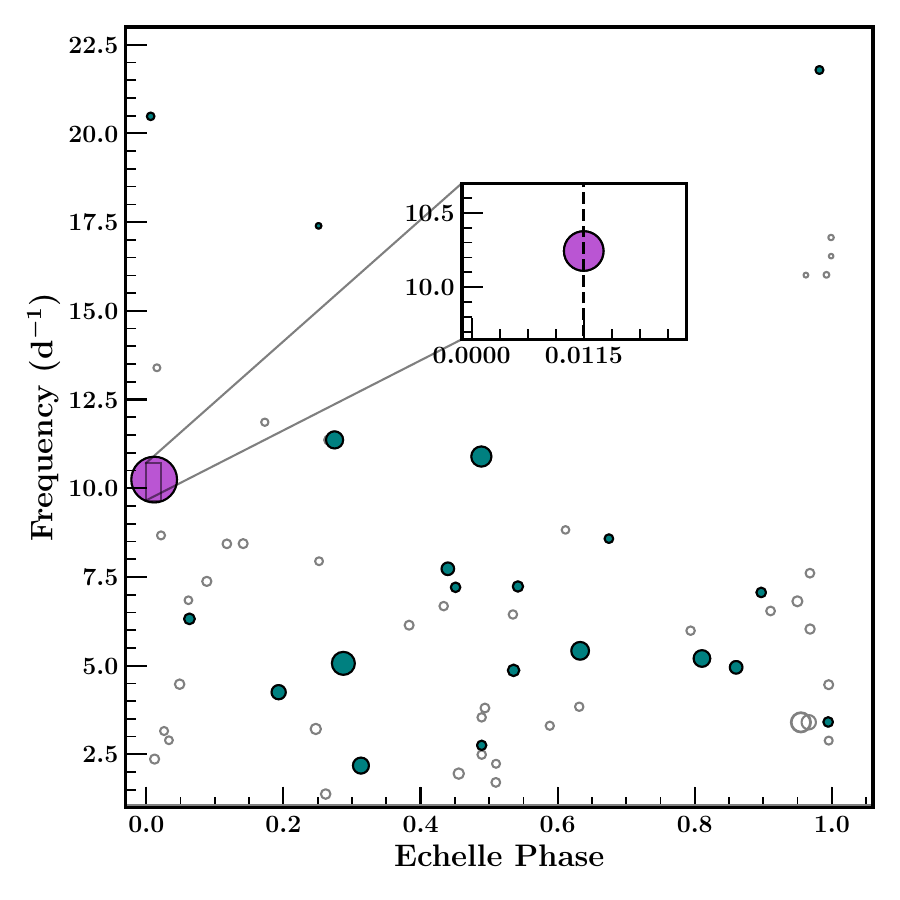} 
    
    \caption{\'{E}chelle diagram of the final set of extracted pulsation frequencies. The \'{e}chelle  phase has been calculated for the orbital frequency. The marker size for the frequencies is proportional to their amplitude. The dominant pulsation frequency is marked in purple, the final set of frequencies is marked in green, and the grey circles are frequencies that were rejected in our analysis.}
        \label{fig:echelle}
\end{figure}

To check for tidal perturbation of the other pulsation frequencies, we created an  \'{e}chelle  diagram where the frequencies were phased with the $f_\mathrm{orb}$ (Fig.\ref{fig:echelle}). We found no multiplets which have similar \'{e}chelle  phases and are separated in frequency by multiples of the $f_\mathrm{orb}$. While our stringent filters rejected the frequencies, we added them to Fig.\ref{fig:echelle} in case they represented any low-amplitude multiplets or unstable multiplets. This was done keeping in mind that the system could be tidally perturbed.
 We did find the rejected frequencies (in grey in Fig.\ref{fig:echelle}) to show multiplets, but they can be due to improper binary modelling (near \'{e}chelle phases 0 or 1) or eclipse mapping of some modes. Unfortunately, with their low amplitudes, it is difficult to conclusively trace their origin.

 \cite{Fuller_2020} showed that the amplitude variations of tidally trapped pulsations are dependent on the pulsating star's $R/R_\mathrm{Roche}$.
 To place \vcep{} in the context of other pulsating close binaries with tidal effects, we consider the Roche-filling factor ($R/R_\mathrm{Roche}$) of the current sample of these systems(see Appendix \ref{app:rochelit}). 

\begin{figure}
    \centering
    \includegraphics[width=\columnwidth]{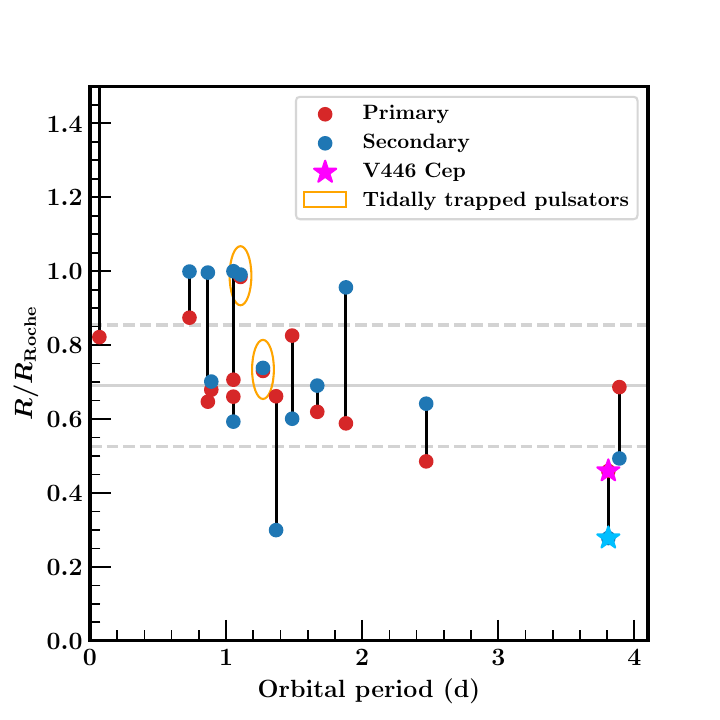}
    \caption{Roche-filling factors ($R/R_\mathrm{Roche}$) versus orbital periods for different tidally perturbed systems with mass and radius measurements in the literature. \vcep\ is marked with a star. The grey lines show the mean and standard deviations for the Roche-filling factors of all pulsators (both secondary and primary, in case the source of pulsation is not identified) in the sample.}
    \label{fig:rocfrac}
\end{figure}

 We used estimates of the orbital periods, masses, and radii of the stars from the literature and then used \phoebe{} to construct a binary model to calculate \texttt{requiv\_max}, which is the $R_\mathrm{Roche}$, for each star. We found that most of the tidally perturbed systems had the pulsating component within 1-$\sigma$ (0.16) of the $R/R_\mathrm{Roche}$ value of 0.69, which is shown in Fig.\ref{fig:rocfrac}. Interestingly, \vcep\ A lies just at the edge of this limit. 

The other stars below \vcep\ A, in Fig.\ref{fig:rocfrac}, are \vcep\ B, and a K star of mass 0.66 $\mathrm{M_{\odot}}$ which is not pulsating \citep{TIC435850195_Jayaraman2024}.  This further contributes to the argument that \vcep\ A is not a tidally perturbed system.

 The dominant pulsation frequency of \vcep\ A is almost an exact harmonic of the $f_\mathrm{orb}$. It is not impossible to have such a coincidence, given that the density of pulsation frequencies is high in \bcep\, stars. But \vcep\, is half way through its tidal synchronisation time and is tens of Myr away from circularisation (see \autoref{sec:evo}) and therefore, will have changes in the orbital frequency. It is possible that its dominant pulsation was or will become an exact harmonic of the orbital frequency.

In Fig.\ref{fig:rocfrac}, the stars labelled as tidally trapped pulsators  (see \citealt{SouthworthBowmanARAAreview} for definitions of different kinds of tidally perturbed pulsators) have two components with large values of $R/R_\mathrm{Roche}$.  On the other hand, the values of $R/R_\mathrm{Roche}$ does not appear to distinguish the tidally tilted pulsators and the non-tilted pulsators. This could hint that other intrinsic factors affect the coupling of tides and pulsations, therefore, it needs further exploration. To calculate $R/R_\mathrm{Roche}$, one needs both mass and radius for the pulsator and, therefore, such pulsators in EBs will prove invaluable.

\section{Third body}
\label{sec:thirdbody}

\vcep\ has long-term proper motion changes, which were visible in {\it Hipparcos} and in \textit{Gaia} second and third data releases (DR2 and DR3; \citealt{dr2binary_Kervella,edr3acceler_Brandt}). In \cite{dr2binary_Kervella}, \vcep\ was identified as an astometric binary but with a primary mass of 6.348 $\mathrm{M_{\odot}}$ and a ``AU-normalised" secondary mass of 363.42 $M_{J}$ or 0.346 $\mathrm{M_{\odot}}$. This estimation uses equation 7 in \cite{dr2binary_Kervella}:
\begin{equation}
    \frac{m_2}{\sqrt{a}} = \sqrt{\frac{m_1}{G}}v_1 ~ ,
\end{equation}
where $m_1$ and $m_2$ denote the primary and secondary masses, respectively, $a$ is the orbital radius, and $v_1$ is the  tangential velocity anomaly calculated in \cite{dr2binary_Kervella}. Using values for $m_1$ and $a$ from our work, we derive $m_2 = 0.149~\mathrm{M_{\odot}}$. Both the estimates are quite small compared to our measured value for $m_2$. This could mean that the orbit (and the companion) identified in \cite{dr2binary_Kervella} is that of a tertiary companion.

\subsection{Eclipse timing}
\label{sec:etv}

In order to confirm the presence of the tertiary component, we applied the timing procedure developed by \citet{timing_Marcadon} to the \textit{TESS} 2-minute cadence data of \vcep{}. 

We adopted the phenomenological model of \citet{2015A&A...584A...8M}, which analytically describes the eclipse profile as
\begin{equation}
    f(t_i,\Theta) = \alpha_0 + \sum^{n_{\rm e}}_{k=1} \alpha_k \, \psi(t_i,T_k,d_k,\Gamma_k,C_k),\label{eq:ecl_profile}
\end{equation}
where $\alpha_0$ is the flux zero-point shift, $n_{\rm e}$ is the number of eclipses during one cycle, and $\alpha_k$ is a scaling coefficient of the eclipse profile function, which is written as
\begin{multline}
    \psi(t_i,T,d,\Gamma,C) = \Bigg\{ 1 + C \bigg( \frac{t_i-T}{d} \bigg)^2 \Bigg\} \\ \times \Bigg\{ 1-\bigg\{ 1-\exp \bigg[ 1-\cosh \bigg( \frac{t_i-T}{d} \bigg) \bigg] \bigg\}^\Gamma \Bigg\}.
\end{multline}
Here, $T$, $d$, $\Gamma$, and $C$ are the time of minimum, the eclipse width, the kurtosis, and the scaling parameter, respectively. For each individual eclipse, the time of minimum is estimated as
\begin{equation}\label{eq:ecl_mod}
    \begin{split}
        T_k & = T_{0,k} + P E + \Delta_k \\
        & = T_{0,k} + P \times {\rm round} \bigg( \frac{t_i-T_{0,k}}{P} \bigg) + \Delta_k,
    \end{split}
\end{equation}
where $T_{0,k}$, $P$, $E$, and $\Delta_k$ are the reference time of eclipse, the orbital period, the epoch, and the observed minus calculated ($O-C$) time difference, respectively. Additionally, the out-of-eclipse variability can be taken into account by including the contribution of the O'Connell and proximity effects in Eq.~(\ref{eq:ecl_profile}), as described by \citet{timing_Marcadon}.

In the first step, we performed an MCMC fit of each sector's light curve assuming that the eclipses are strictly periodic (i.e., $\Delta_k = 0$ for each eclipse). For each sector, we obtained a set of parameters that best fit the corresponding light curve from a chain of $10^5$ iterations. Given that the light curve of \vcep\ is stable with time (no evidence of activity-induced
variability is found), we adopted the set with the smallest value of normalised $\chi^2$, among the 10 single-sector light curves, as reference. In the second step, we generated a chain of $10^4$ iterations for each available sector, leaving $\Delta_k$ as a free parameter and fixing the other parameters to their best-fitting values. Finally, we calculated the times of minima from the best-fitting values of the parameter $\Delta_k$ using equation~(\ref{eq:ecl_mod}) and removed those occurring during data gaps. The times of minima derived from our fitting procedure are listed in Table~\ref{tab:ecl_times} in Appendix~\ref{app:ecl_times}, along with the errors representing 1$\sigma$ percentiles from the MCMC posteriors. The minima fitting does not account for errors due to intrinsic pulsations in the system. So there might be an underestimation of errors. 

\subsection{Period Changes}
\label{sec:pultiming}

The amplitude of the dominant pulsation, $f_1$, is around 2.5~per~cent of the primary eclipse, 6.7~per~cent of the secondary eclipse, and 400~per~cent of the frequency with the second highest amplitude. 

To complement the eclipse timing method for detecting a tertiary component, we investigated the period changes (PC) of the dominant pulsation frequency of \vcep\ to search for signatures of the light-travel time effect \citep[LTTE,;][]{Irwin1952ApJ, Irwin1959AJ}. Our objective was to apply a Fourier-based, modified Hertzsprung method \citep[see e.g.][]{Hajdu2021ApJ, Rodriguez-Segovia2022MNRAS} to derive the $O-C$ variations. The detailed methodology is outlined in \citet{timing_rathour, Rathour2025A&A}. Each \textit{TESS} sector, with carefully subtracted eclipse signatures, served as the starting point of the analysis. 
 We removed all significant frequencies except the dominant mode at $f_{1} \simeq 10.24324~\mathrm{d}^{-1}$. 
 Once the overall data were cleaned for additional signals, we investigated for PC behaviour of the dominant pulsation frequency $f_{1}$. Unlike the eclipse timing technique, which measures the timing of individual eclipses, the Hertzsprung method tracks phase shifts of the pulsation light curve relative to a reference template that is constructed from a Fourier fit to the light curve. The comparison of any given observed light curve segment with the template results in a phase shift over time, hence providing a direct measure of a period change (see e.g. \citealt{Bowman2016, Bowman2021}). Uncertainties for individual $O-C$ points were estimated using a bootstrap resampling approach \citep[see][]{Hajdu2021ApJ}.

\begin{figure}
    \centering
    \includegraphics[width=\columnwidth]{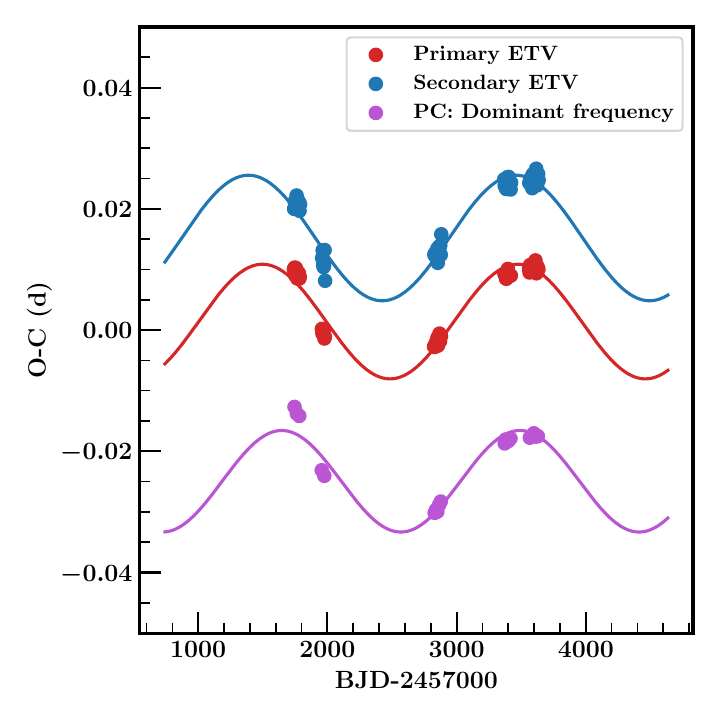}
    \caption{Timing variations for primary eclipses (in red),  secondary eclipses (in blue), and the period change (PC) of the dominant pulsation frequency (in purple; shifted for clarity). The lines show sinusoidal fits with a period corresponding to 2310 days. }
    \label{fig:etv}
\end{figure}

The resulting $O-C$ diagram derived from period change analysis and the ETVs, corresponding to both the primary and secondary eclipses, is shown in Fig.\ref{fig:etv}. It shows sinusoidal variations with a period above 2000~d,  with the amplitudes of the order of 0.010~d. This consistent sinusoidal signal indicated the presence of a tertiary stellar companion.  We rule out any evolutionary change as the timescale is short. We can rule out apsidal motion and mass transfer, because we do not have anti-correlated or inconsistent signals from the primary and secondary ETVs, and PC. The coherent time variations from the two different methods also mean that the source of eclipses and the pulsation is the same. Therefore, we confirm that the pulsation originates from the binary system and is not a result of contamination in the large \textit{TESS} pixels.

\subsection{Combining radial velocities}
Using traditional RV extraction methods, we found it difficult to extract the faint secondary. But we were able to extract the primary RVs both using broadening functions (BF; \citealt{Rucinski_BF_1992}) and two-dimensional cross-correlation (\textsc{todcor}; \citealt{Zucker_TODCOR_1994}) methods. These RVs were similar to the RV time series we extracted using \spd{}. We also fit the RVs from the BF method with a binary model and used the residuals to search for the tertiary companion. 

To get a better coverage of the orbital phase of the outer orbit, we combined these RVs with the pulsation-derived timing variations (i.e., $O-C$) converted into an equivalent RV curve. We transformed the $O-C$ points into RV using:

\begin{equation} 
\label{eq:oc_to_rv}
v_{\mathrm{rad}} = c\frac{{\rm d} \tau}{{\rm d} t}
\end{equation}

\noindent where $c$ is the speed of light and $\tau = (O-C) \times 86400.0$ denotes the LTTE delay in seconds. The time derivative, ${\rm d}\tau/{\rm d}t$, was evaluated numerically using a finite-difference gradient method.

\begin{figure}
    \centering
    \includegraphics[width=\columnwidth]{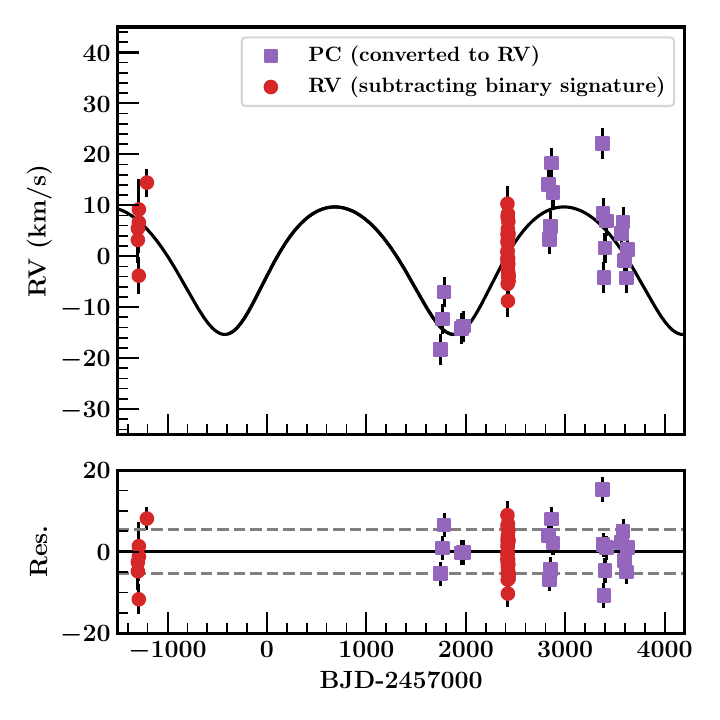}
    \caption{ Joint RV-PC fit for the tertiary (in black). The RVs (in red circles) and PCs (in purple squares) are spread over a period of around 5000 d while the tertiary is in an orbit of 2303~d. The bottom panel shows the residuals after the fit. The dashed lines show the {\it rms} of the scatter. The error bars on the points represent weighted errors as described in the text.}
    \label{fig:jointpcrv}
\end{figure}

Using this joint RV-PC dataset, we used \textsc{v2fit} \citep{Konacki_2010} to fit for the tertiary star. Unfortunately, the scatter of the RVs was large, and the orbital phase coverage was poor. But together with the converted PCs, they help constrain the outer orbital period ($P_2$). 
Notably, the RV converted from PCs has a much larger scatter than their individual errors; therefore, we included a jitter
term (added in quadrature to individual errors), which are also treated as a free parameter in {\sc v2fit}. We found the jitter to be 2.986~km~s$^{-1}$. All RVs were weighted according to their uncertainties, i.e.,  direct RVs from their errors, and converted-PC RVs from their final errors, which include the jitter. We found an eccentric orbit with eccentricity $e_2=0.15$. We found large errors on $P_2$ and time of periastron passage ($T_{0,2}$), expected because of poor orbital phase coverage. Nonetheless, we found the projected mass of the tertiary ($M_{3}\sin^3{i_{2}} $) to be $4.10 \pm 0.32$~$\mathrm{M_{\odot}}$. The resulting fitting parameters are in \autoref{tab:jointfit}. The orbital fit is shown in Fig.\ref{fig:jointpcrv}. 

\begin{table}
    \centering
    \begin{minipage}{45mm}
    \caption{ Tertiary parameters estimated from \textsc{v2fit} and \textsc{gaiamock}.}
 \label{tab:jointfit}
        \renewcommand{\arraystretch}{1.0}
    \begin{tabular}{cc}
    \hline
    \hline
          Parameter & Value\\
    \hline
          $P_2$ [d] & $2303\pm  69$ \\  
          $T_{0,2}$ [d] &  $2456629 \pm 225 $\\
          $e_2$ & $0.150   \pm   0.098$  \\ 
           $i_{2}$ [deg]& $88 \pm 7$  \\
           $\Omega_{2}$ [deg]& $173 \pm 105$\\
           $f_{2}$ &  $0.166\pm 0.014$ \\
          $a\sin{i_{2}}$ [AU]& $7.886 \pm 0.243$ \\
          $K_3$ [km\,s$^{-1}$] & $ 12.5 \pm 1.7$ \\
          $M_{3}\sin^3{i_{2}} $ [$\mathrm{M_{\odot}}$] & $4.10 \pm 0.32$ \\
         Angular size [mas] & $17.5 \pm 3.0$ \\
    \hline
    \end{tabular} 
    \end{minipage}
\end{table}

\subsection{Comparison with astrometric simulations}

To better understand the configuration of the companion, we used \mbox{\textsc{gaiamock}\footnote{https://github.com/kareemelbadry/gaiamock}} \citep{Badry_2024} to simulate the system with different configurations of the tertiary orbit.  \textsc{gaiamock} also calculates the \textit{Gaia} re-normalised unit weight error (RUWE) for a certain configuration of orbital parameters like the masses, period, sky-inclination ($i_{2}$), longitude of the ascending node ($\Omega_{2}$) of the orbit of a system, and flux-ratio of the system ($f_{2}$). In our case, we treated the eclipsing binary as a single star and the whole system as a binary. Therefore, the \textsc{gaiamock} setup was initialized as a $12.34~\mathrm{M_{\odot}}$ primary and a secondary star with a mass of $4.1/\sin^3{i_{2}}$~$\mathrm{M_{\odot}}$. 

\begin{figure}
    \centering
    \includegraphics[width=\columnwidth]{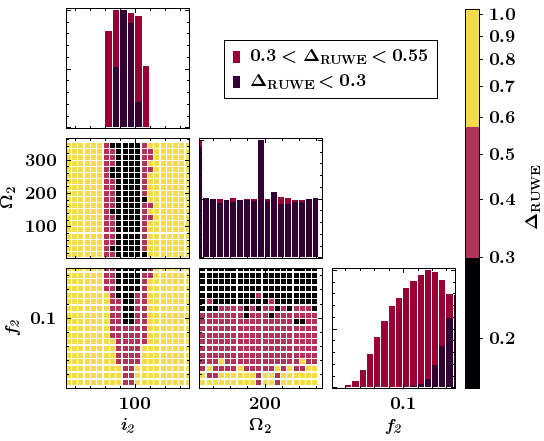}
    \caption{Distribution of $\Delta_\mathrm{RUWE}$ for \textsc{gaiamock} models with different values of $i_{2}$, $f_{2}$, and $\Omega_{2}$. The models in dark violet have $\Delta _\mathrm{RUWE} < 0.3$ and represent the physically feasible models.}
    \label{fig:gaiamock}
\end{figure}

We simulated different models of this binary with different combinations of  $i_{2}$, $f_{2}$, and $\Omega_{2}$ and kept fixed the parallax, orbital period, and eccentricity. For these models, we calculated $\Delta_\mathrm{RUWE}$, which is defined as:
\begin{equation}
    \Delta_\mathrm{RUWE}=|\mathrm{RUWE}_{Gaia}- \mathrm{RUWE_{mod}}|
\end{equation}
where $\mathrm{RUWE_{mod}}$ is calculated from \textsc{gaiamock} and $\mathrm{RUWE}_{Gaia}$  is the value from \textit{Gaia} DR3. We found that all the modelled configurations had a positive $\Delta_\mathrm{RUWE}$.
From this grid of models, we picked models that had the lowest $\Delta_\mathrm{RUWE}$. Then, we estimated orbital parameters from the mean and standard deviation of the distributions of  $i_{2}$, $f_{2}$, and $\Omega_{2}$ for the models with $\Delta_\mathrm{RUWE}$ less than 0.3 (Fig.\ref{fig:gaiamock}). This limit was  estimated from a sample of close triple systems with known orbital parameters (see Appendix \ref{appendix:chtruwe}). There were 3076 such models, compared to the total sample of 128\,000, which we identify as the most probable solutions.  We find that these models had $i_{2}$ of $88 \pm 7 ^{\circ}$ and $\mathrm{f}_2$ around 0.166$\pm$0.014. There was no obvious preference for $\Omega_{2}$ but slightly more models favoured 0$^{\circ}$ and 180$^{\circ}$. 

The most probable configuration gives us a tertiary with a mass of  $4.11 \pm 0.32 \mathrm{M}_{\odot}$. An $i_{2}$ around $88^{\circ}$  means that the system is co-planar and therefore there is a possibility of observing extra eclipses in the system. With the estimated $T_{0,2}$, we expected extra eclipses in {\it TESS} sectors 83 to 85. Unfortunately, we found no such eclipses. We also calculated the angular size of the system with errors calculated assuming an error of 150 pc in the {\it Gaia} distances. The orbital parameters of the tertiary are listed in \autoref{tab:jointfit}. 

We also note that a {\sc gaiamock} model with no tertiary gives a $\mathrm{RUWE_{mod}}$~=1.01 and therefore is within the our limit for acceptable solutions. But there is no indication that the ETVs and period changes could arise from any source other than LTTE due to the teritary. This calls for caution when using the orbital parameters estimated from the astrometric simulations.

\section{Evolution and formation}
\label{sec:evo}

The inner binary of \vcep\, has a low mass-ratio. If the system was formed this way, it presents an interesting case to study the formation of close binaries in massive stars. Another possible scenario is that mass transfer in the past changed the configuration of the system from a higher mass ratio to a smaller mass ratio. We found no evidence of mass transfer in the abundance patterns. To shed more light, we decided to check if the system is co-evolving. The parameters of the newly found tertiary companion also need to be taken into consideration to get the whole picture of the evolutionary state of the system. 

We used the spectral energy distribution (SED) for expected multiple-star configurations to constrain the stellar parameters of the companion. For this, we used the Python routine \textsc{SEDfit}\footnote{\url{https://github.com/mkounkel/SEDFit}}. \textsc{SEDfit} allows fitting a multiple-star SED, assuming a single metallicity for the whole system, by querying archival photometric fluxes in ultraviolet, visible, and infrared bands. The individual SED for each component is generated using Kurucz model atmospheres \citep{1992IAUS..149..225K}. We fit a combined SED to three possible cases for the system: (i) a binary with a compact object (no additional stellar flux from the companion); (ii) a triple system with a main-sequence star (where we expect flux from the third star); and (iii) a quadruple system where the EB is orbited by another binary (consisting of similar stars). We fixed the radius, $\log{g}$, $T_{\rm eff}$ of the binary stars from the light curve modelling and spectroscopic analysis. The metallicity was set to zero with a range of variation set from $-$0.05 dex to 0.05 dex. For the companion(s), we set an initial search range of 0.5 $\mathrm{R}_{\odot}$  to 3 $\mathrm{R}_{\odot}$ in radius, 3.5 dex to 4.6 dex in $\log{g}$, and 2000 K to 10,000 K in temperature. We then changed the ranges separately for case-(ii) and case-(iii) to get a best fitting solution which gave the mass/combined mass\footnote{The code fits for radius and $\log{g}$. We calculate the mass using $  M=  \frac{R^2}{A_{c}^2} 10^{\log (g)} $, where $A_c \equiv \sqrt{G M_\odot}/R_\odot $.  \\} close to $4.11 \pm 0.32 \, \mathrm{M}_{\odot}$. The final fits are shown in Fig.\ref{fig:sed}. After we obtained SED parameters from the fitting, we used the expected mass, radius, and temperature of the companion, and the EB, to estimate the age. We did this by fitting isochrones from MESA Isochrones and Stellar Tracks (MIST; \citealt{mistc2016,mistd2016}).  We avoid any quantitative constraints from the third light. This is because the third light is affected by light curve detrending and the removal of pulsation signature. We still consider the third light for a qualitative discussion. With these constraints, we discuss the possible formation and evolution of the system below. 

\subsection{Tertiary is a compact object}

We find that case-(i)  (left panel of Fig.\ref{fig:sed}) 
keeps the system at a distance of 731.6 pc, compared to the reported \textit{Gaia} distance of 900 pc but gives us a plausible SED fit for two stars and hence makes the tertiary a compact object.
The mass of the tertiary is large compared to the Tolman-Oppenheimer-Volkoff (TOV) limit for the maximum mass of a neutron star. Therefore, the tertiary could be a black hole or a binary containing neutron stars. A black hole tertiary with mass of 4.11~M$_{\odot}$ implies that it evolved from a 20-30~M$_{\odot}$ main sequence star in about 10-14 Myrs \citep{1999ApJ...522..413F,2019Natur.575..618L}. We see third-light in the light curve solution, which acts against accepting this scenario, but that could also arise from other structures within the system, for example accretion discs. It also plausible that the third light is from a background star as \vcep\, is close to the galactic plane. We also note that there were no signatures of a supernova explosion in the atmospheric abundances of the primary. But it is also possible that the supernova event was quick enough to not affect the chemical abundances of the primary. In another possible scenario, the tertiary could have been formed elsewhere and was captured dynamically in the current system. 

The best-fitting age of the binary from the isochrone fitting is 14.5 Myr (Fig.\ref{fig:isochrones}; left), which is almost twice the median age of a sample of \bcep{} stars observed by \textit{TESS}, which is 7 Myr \citep{Fritzewski_2025}. But the age is small compared to the theoretical timescales of the tidal evolution of the system. We use the code {\sc jktabsdim}\footnote{\url{https://www.astro.keele.ac.uk/jkt/codes/jktabsdim.html}} (based on \citealt{Zahn_1975,Zahn_1977}) to calculate theoretical synchronisation and circularisation times. We found that the synchronisation and circularisation timescales are 31 Myrs and 114 Myrs, respectively. This explains the super-synchronous rotation of the components. The age of the system also supports the possibility of a compact object as the tertiary.

\subsection{Tertiary is a main-sequence star}
Case-(ii) SED fitting (centre panel of Fig.\ref{fig:sed}) shows that the tertiary flux is higher than the secondary flux, at all wavelengths. We cannot fit a single isochrone to such a tertiary, and hence, it could not be co-evolving. Instead, we find that such a companion is only possible if it formed later and is metal-rich compared to the inner binary (centre panel of Fig.\ref{fig:isochrones}). In this case, the tertiary is less massive than the calculated minimum mass. The infrared excess in the SED residuals is also unexplained. Furthermore, in such a case, we would also expect to see the lines of the tertiary in the optical spectra. 
The above reasons make this case the least favourable of all three cases.

\subsection{Tertiary is a close binary}
\begin{figure*}
    \includegraphics[width=0.33\textwidth]{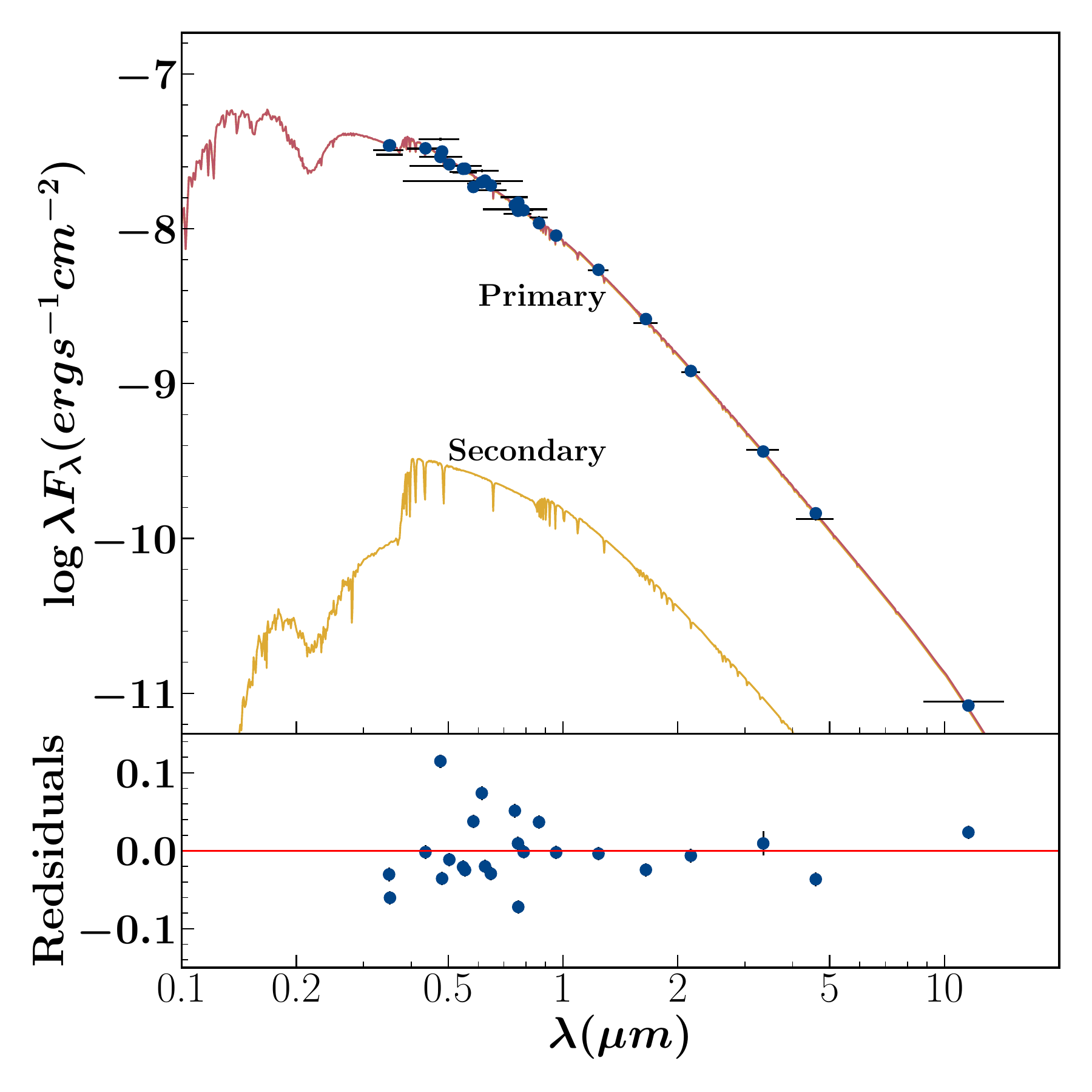} 
  \includegraphics[width=0.33\textwidth]{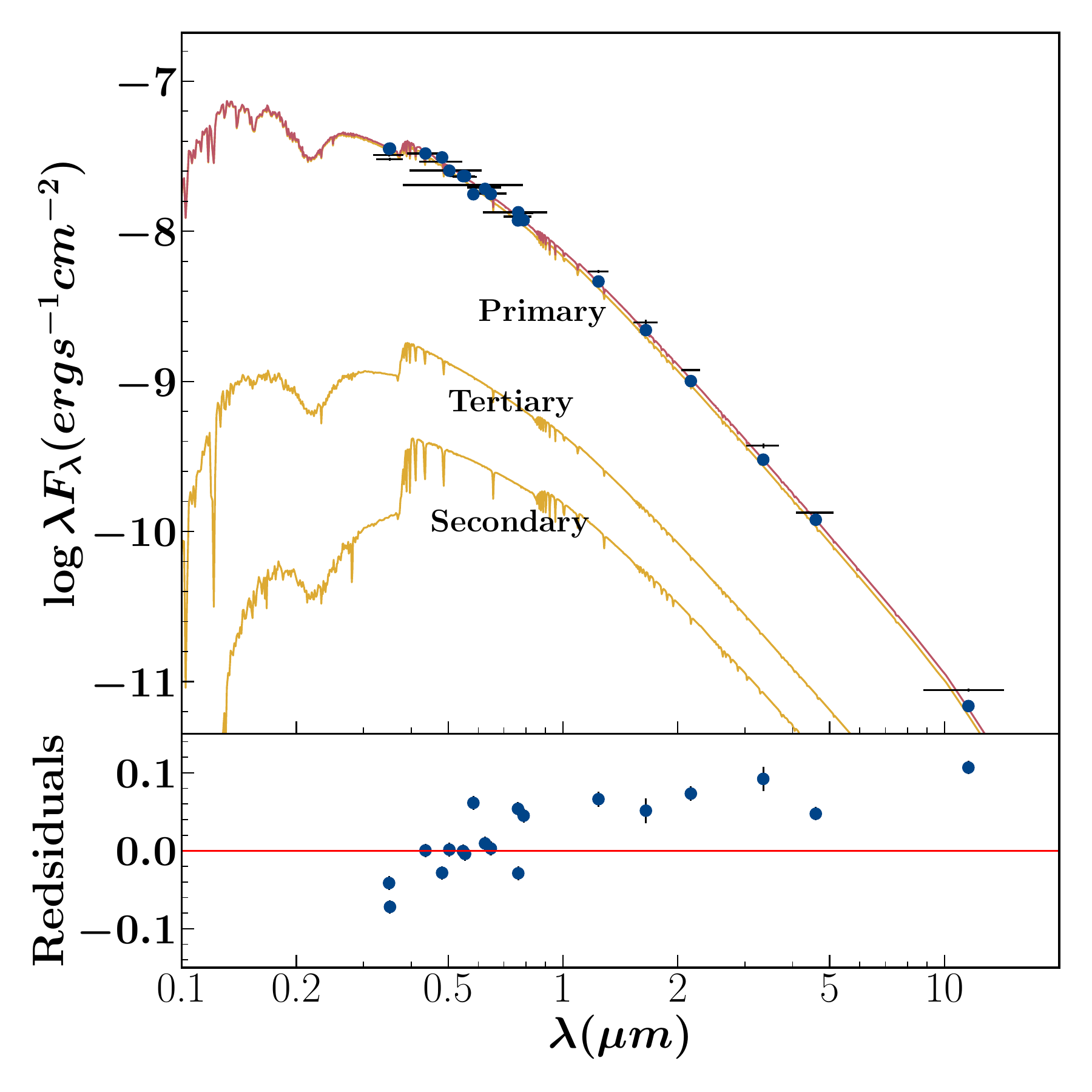} 
  \includegraphics[width=0.33\textwidth]{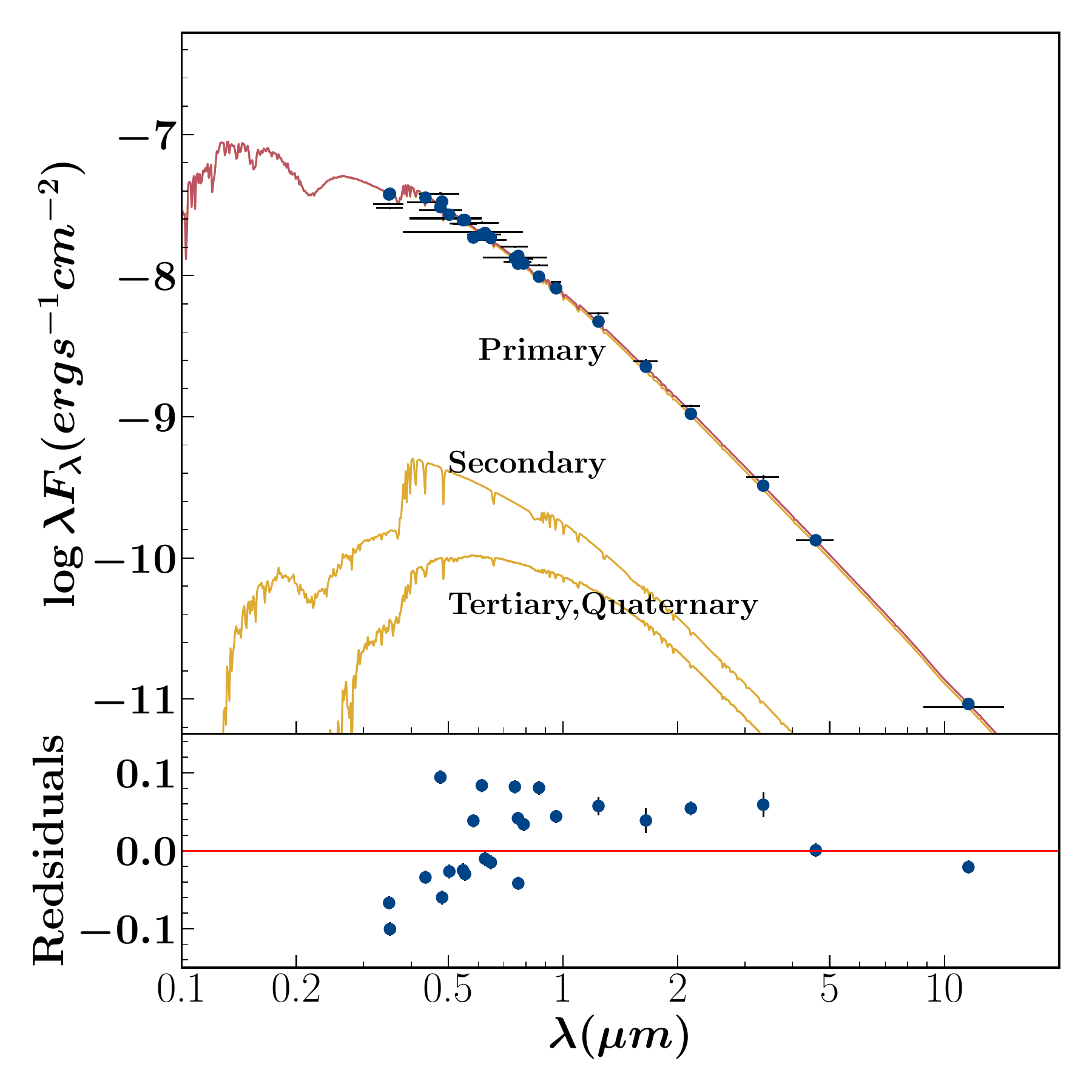}     
  \caption{Best-fitting Kurucz SED for the binary+compact object case (left), main sequence triple case (centre), and the quadruple case where the third and fourth star have the same SED (right). The yellow lines show the component SEDs, while the red line shows the combination SED. The primary has the maximum flux contribution so its SED is close to the total SED. The dots represent the modelled fluxes and the crosses represent observed fluxes.  The bottom panels show residuals for different photometric observations. The errors on the photometry are underestimated as they do not account for eclipses and pulsations, which significantly affect the total observed flux of the system.}
  \label{fig:sed}
\end{figure*}

\begin{figure*}
    \centering
\includegraphics[width=0.33\textwidth]{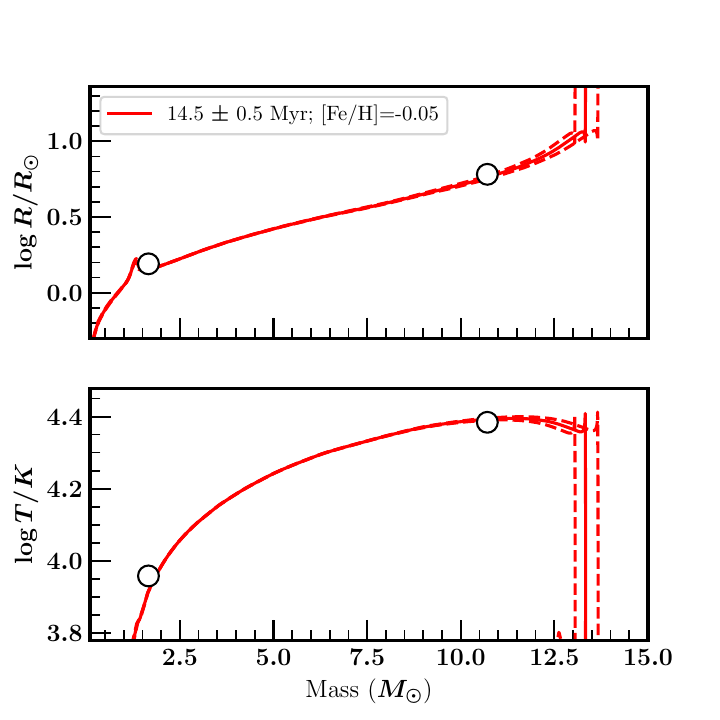}
\includegraphics[width=0.33\textwidth]{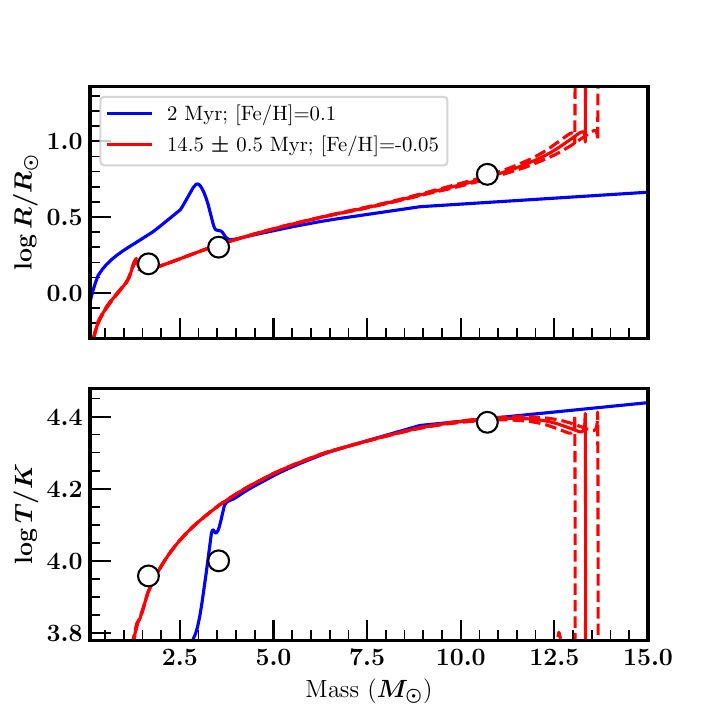}
\includegraphics[width=0.33\textwidth]{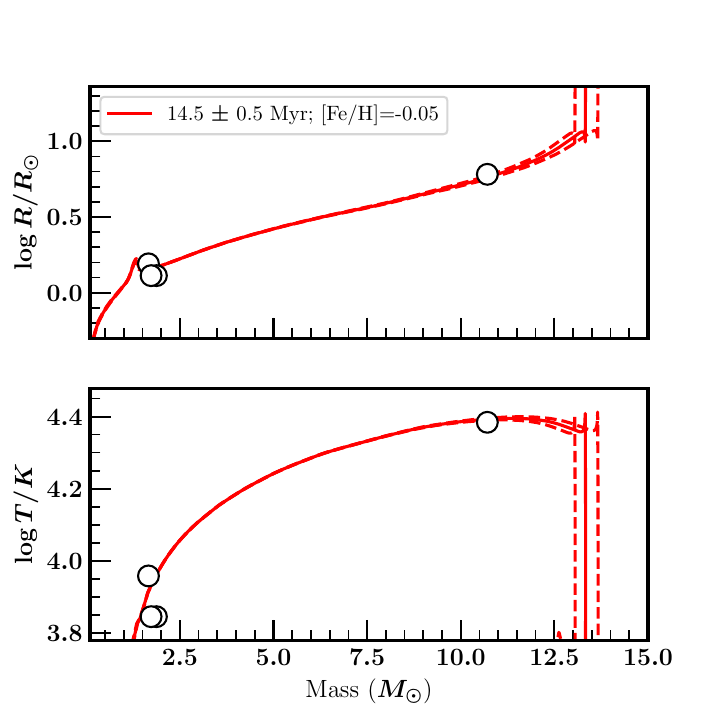}
    \caption{Best-fitting MIST isochrones for the binary+compact object case (left), main sequence triple case (centre), and the quadruple case (right). The red isochrone shows the best fit for the pulsating binary system, while the blue line shows an isochrone of 2 Myr. The stars are represented as hollow circles. Their error bars are smaller than their markers.}
    \label{fig:isochrones}
\end{figure*}

Case-(iii) (right panel of Fig.\ref{fig:sed}) is the most simplistic case. This case represents a scenario where a close binary, with main-sequence stars, is the companion. This gives rise to the expected third light, explains the mass calculated from the RV-PC fitting, and has the same metallicity and age as that of the EB (right panel of Fig.\ref{fig:isochrones}). The individual contribution of the stars to the total flux is lower than that of the secondary and this would explain why we do not see additional lines in the spectra. This scenario makes \vcep\ a co-evolving hierarchical quadruple where all the stars were formed in situ.

\begin{table*}
    \centering
        \caption{Parameters obtained from SED fitting for the three cases discussed in \autoref{sec:evo}.}
    \begin{tabular}{cccc}
    \hline
Parameters  & Binary (tertiary is compact object) & Triple  & Quadruple\\
    \hline
        $A_{v}$& 1.00& 0.85& 0.83\\
 $d [{\rm pc}]$& 731.6 & 800.0& 750.0\\
 R [$\mathrm{R}_{\odot}$]& [5.9 1.5]& [5.9, 1.5,  2.  ]& [5.9, 1.5,  1.3,  1.3 ]\\
 $\log{g}$& [3.9, 4.3]& [3.9, 4.3,  4.4 ]& [3.9, 4.3,  4.5, 4.5]\\
 $T_\mathrm{eff} [{\rm K}]$& [24230,  9080]& [24230,  9500, 10000]& [24220,  9480,  7000,  7000]\\
 $[{\rm Fe}/{\rm H}]$& -0.05& 0.05& -0.05\\
 Total companion mass [$\mathrm{M_{\odot}}$] & - & 3.5 & 4.4 \\

 $\chi^{2}$& 23& 31.0& 38.0\\
 \hline
    \end{tabular}

    \label{tab:sed_fit}
\end{table*}

\section{Conclusions}

We analysed the pulsating EB \vcep\ with ten sectors of \textit{TESS} photometry and 72 HERMES spectra. Using detailed light curve modelling and orbital analysis, we found the stellar and orbital parameters of the system. The system contains a 3.808567 d binary with a low mass-ratio of 0.155. We also constrained the atmospheric parameters of the stars from disentangled spectra. The primary was found to be the hotter component with a temperature of 24220 K compared to the 9080 K secondary. 

Using the residuals of the binary light curve modelling, we extracted the pulsation frequencies. 
From the masses and temperatures of the companions, we attribute the pulsations to the primary component.  The dominant pulsation frequency is a near-harmonic of the orbital frequency. We found no multiplets or significant phase and amplitude variations over the orbital period to suggest that it is a tidally perturbed mode.   More spectroscopic observations, spanning a few orbital periods, would be helpful in identifying the mode(s).

We also found significant ETVs for the primary and secondary eclipses. Using pulsation timing of the dominant pulsation frequency, we found a consistent period variation of around 2408 d with an amplitude of 0.15 d. We combined RV and period changes of the dominant pulsation to find that the newly discovered companion has a minimum mass of 4.10$\pm$0.32~$\mathrm{M_{\odot}}$. We also used astrometric models and \textit{Gaia} astrometric parameters to get orbital parameters of the tertiary orbit. 

We explored the formation and evolution of the system with SED modelling and isochrone fitting. We found three cases, out of which two possibilities exist. The tertiary is either a compact object or a binary with main-sequence stars. In the compact-object case, the mass constraints are satisfied by either a single black hole or a double neutron-star binary. If the tertiary is a close main-sequence binary, the stars are co-evolving along with the inner eclipsing binary.

Identification of the true nature of the tertiary needs more observations. New radial velocities, during orbital phases close to the maximum amplitude of the tertiary orbit, will be helpful. The size of the tertiary orbit also allows for an interferometric determination of a more accurate and three-dimensional architecture of the orbit. However, the stars in the EB are well characterised, especially the \bcep\ primary component. It offers a good set of observables for detailed evolutionary modelling which is necessary since we see tentative evidence of tides impacting the pulsations. 

\section*{Acknowledgements}
The authors thank the anonymous referee for their suggestions which helped in improving the manuscript. 

A.\ Moharana thanks Jeppe Sinkbaek Thomsen for comprehensive discussions on pulsations and light curve modelling. A. Moharana also thanks Dr. Kareem El-Badry for useful discussions regarding \textsc{gaiamock}.  The authors also thank Dr. Raphael Hirschi, Dr. Pierre Maxted and Dr. Barry Smalley for useful discussions.

A.\ Moharana and JS acknowledge support from the Science and Technology Facilities Council (STFC) under grant number ST/Y002563/1.

A.\ Miszuda acknowledges the support by the Polish National Science Centre (NCN), grant number 2021/43/B/ST9/02972.

RSR acknowledges support from the French Agence Nationale de la Recherche (ANR) under grant ANR-23-CE31-0009-01 (UnlockPFactor), and from the NCN through the Sonata BIS 2018/30/E/ST9/00598 and the OPUS grant 2024/53/B/ST9/02630.

KGH acknowledges the support of the NCN through the grant 2023/49/B/ST9/01671.

DMB gratefully acknowledges UK Research and Innovation (UKRI) in the form of a Frontier Research grant under the UK government's ERC Horizon Europe funding guarantee (SYMPHONY; PI Bowman; grant number: EP/Y031059/1), and a Royal Society University Research Fellowship (PI Bowman; grant number: URF\textbackslash R1\textbackslash 231631).

This work used the greenHPC facility at Keele University, which is supported by the Wolfson Foundation's \href{https://www.wolfson.org.uk/funding/funding-for-places/}{Funding for places} grant.

 Based on observations made with the Mercator Telescope, operated on the island of La Palma by the Flemish Community, at the Spanish Observatorio del Roque de los Muchachos of the Instituto de Astrofísica de Canarias. Based on observations obtained with the HERMES spectrograph, which is supported by the Research Foundation - Flanders (FWO), Belgium, the Research Council of KU Leuven, Belgium, the Fonds National de la Recherche Scientifique (F.R.S.-FNRS), Belgium, the Royal Observatory of Belgium, the Observatoire de Genève, Switzerland and the Thüringer Landessternwarte Tautenburg, Germany.

This paper includes data collected with the \textit{TESS} mission, obtained from the Mikulski Archive for Space Telescopes (MAST) at the Space Telescope Science Institute (STScI). Funding for the \textit{TESS} mission is provided by the NASA Explorer Program. STScI is operated by the Association of Universities for Research in Astronomy, Inc., under NASA contract NAS 5–26555.
This work has made use of data from the European Space Agency (ESA) mission {\it Gaia} (\url{https://www.cosmos.esa.int/gaia}), processed by the {\it Gaia}
Data Processing and Analysis Consortium (DPAC, \url{https://www.cosmos.esa.int/web/gaia/dpac/consortium}). Funding for the DPAC has been provided by national institutions, in particular the institutions participating in the {\it Gaia} Multilateral Agreement. This work made use of TOPCAT and the STILTS package \citep{topcat,stilts}.
This research has made use of the SIMBAD database \citep{simbad_2000},
operated at CDS, Strasbourg, France. This research has made use of the Astrophysics Data System, funded by NASA under Cooperative Agreement 80NSSC25M7105.

This is a pre-copyedited, author-produced PDF of an article accepted for publication in MNRAS following peer review. The version of record is available online at: \url{https://doi.org/10.1093/mnras/stag772}

\section*{Data Availability}

The \textit{TESS} data used in this article are available in the MAST data archive (\url{https://mast.stsci.edu/portal/Mashup/Clients/Mast/Portal.html}). The HERMES spectra used in the article are available at:  \url{https://doi.org/10.5281/zenodo.17513474}. The ETV measurements are available in the online supplementary material corresponding to the article.



\bibliographystyle{mnras}
\bibliography{v446cep} 



\appendix

\section{diagnostics of \phoebe\, modelling }
\begin{figure*}
    \includegraphics[width=\textwidth]{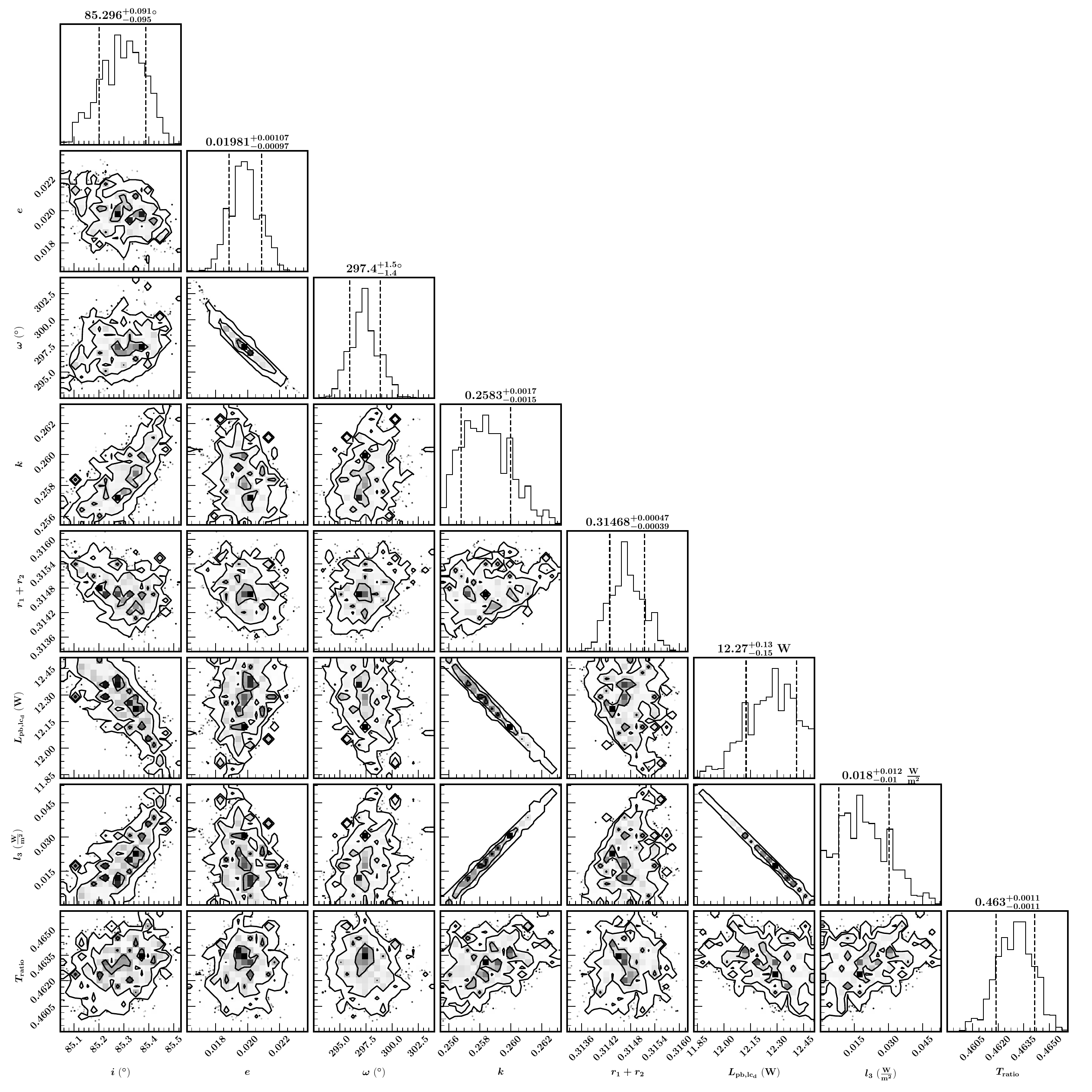}
    \caption{Corner plots for the \phoebe\, MCMC sampling of stellar and orbital parameters of the EB in \vcep.}
    \label{fig:corner}
\end{figure*}

\begin{figure}
    \centering
    \includegraphics[width=\columnwidth]{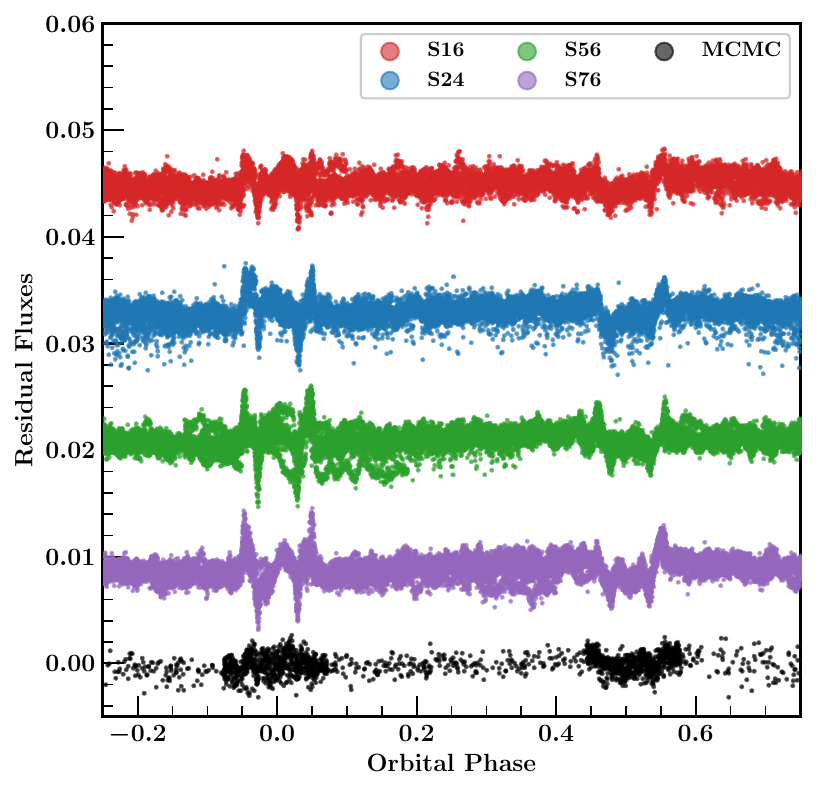}
    \caption{Light curve residuals after subtraction of \phoebe\, models. }
    \label{fig:rescomp}
\end{figure}

\begin{figure}
    \centering
    \includegraphics[width=\columnwidth]{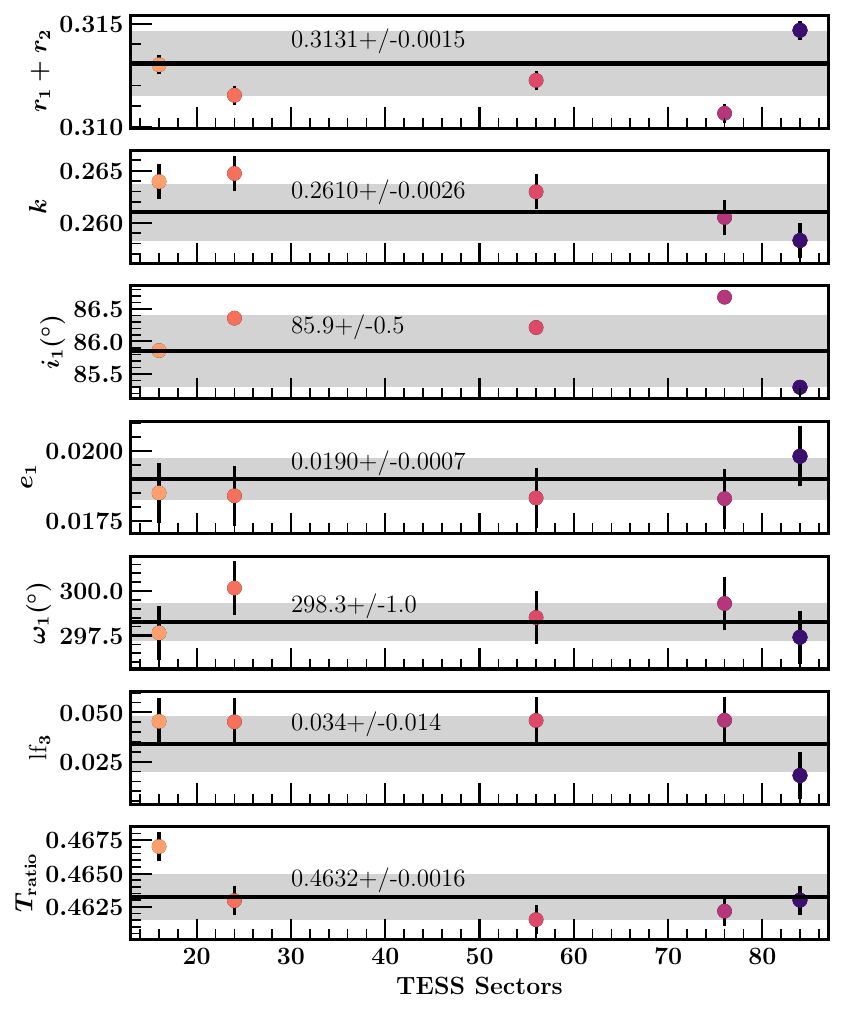}
    \caption{Variation of parameter values over different {\it TESS} sectors. The black points represent the MCMC estimate. The black line in every panel is the adopted value and the grey shaded region represents the range of the adopted errors of the corresponding parameters. Top to bottom: sum of fractional radii, radius ratio, binary inclination, eccentricity, argument of periastron, third light fraction, and ratio of effective temperatures. }
    \label{fig:paramvar}
\end{figure}

The corner plot for the adopted MCMC chain is shown in Fig.\ref{fig:corner}
The final residuals after the Nelder-Mean optimisation for different sectors are given in Fig.\ref{fig:rescomp}. The MCMC residuals are shown in black for comparison. The parameter consistency is shown in Fig.\ref{fig:paramvar}.
\label{appendix:mcmc}

\section{Broadening functions}
\label{appendix:BF}
\begin{figure}
    \centering
    \includegraphics[width=\columnwidth]{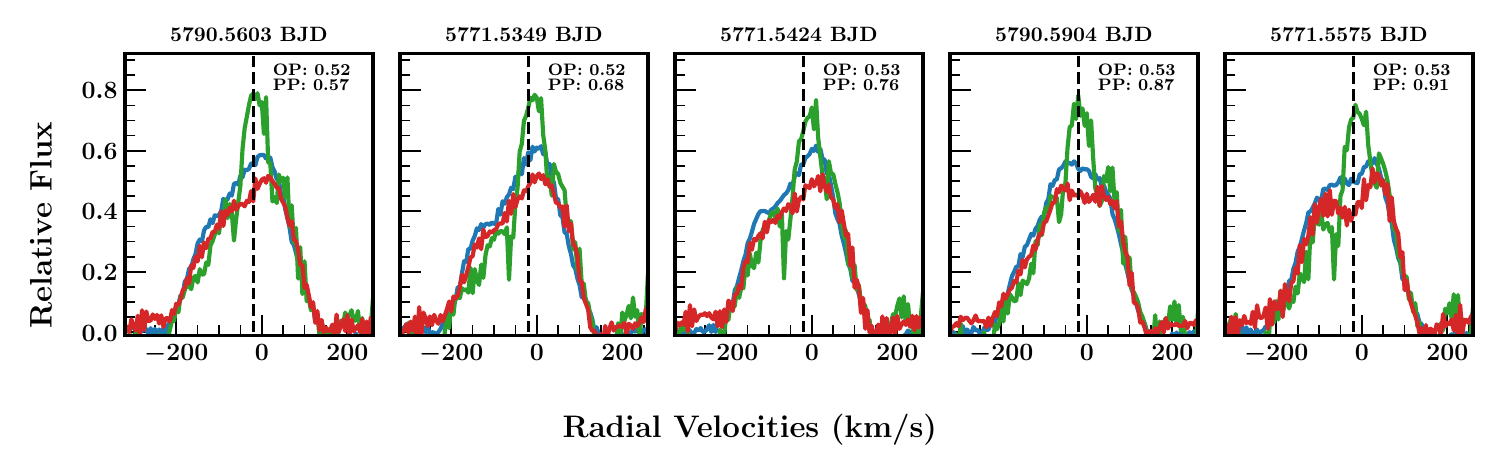}
        \includegraphics[width=\columnwidth]{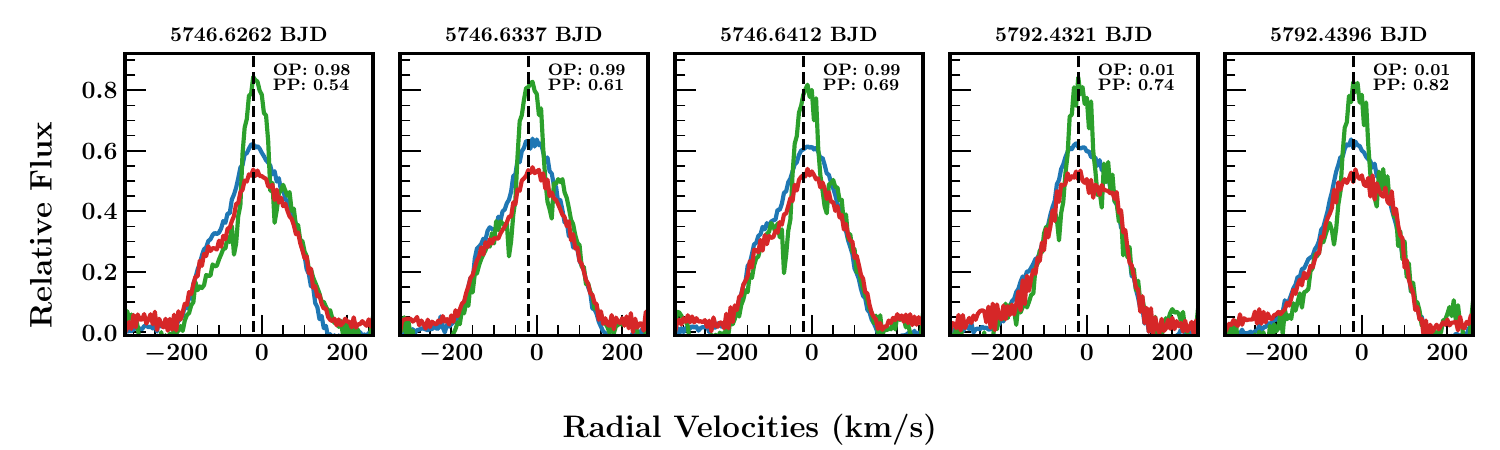}
    \caption{BF variation during secondary (total) eclipse (top) and primary eclipse (bottom). BF in blue, green, and red are extracted from spectra in regions of 4000-5000 A$^\circ$, 5000-6000 A$^\circ$, and 6000-7000 A$^\circ$, respectively. The dashed black line is the gamma velocity.}
    \label{fig:BF_ecl}
\end{figure}

The broadening function (BF; \citealt{Rucinski_BF_1992}) has the capability to visualise intrinsic stellar effects like rotational broadening, spots, and pulsations \citep{Moharana_ST3}. To spot similar variations in the \bcep{} component of \vcep, we calculated BF using the HERMES spectra. The BF was calculated using the algorithm described in \citet{BFSVD_Rucinski}. We modified a single-order BF code, \textsc{bf-rvplotter}\footnote
{\url{https://github.com/mrawls/BF-rvplotter}}, to calculate multi-order BFs. The BF was calculated in a wavelength range of 4000-5000 \AA, 5000-6000 \AA, and 6000-7000 \AA. As our template, We used a synthetic spectrum generated with the code \textsc{spectrum} using model atmospheres from \textsc{ATLAS9}, abundances from \cite{2009ARA&A..47..481A}, and line lists from \cite{2022A&A...666A.121R}. The template created had a temperature of 18800 K, $\log{g}$ of 3.5 dex, [M/H] of 0.5 dex with  projected rotational velocity ($v\mathrm{sin}i$) of 1 $\mathrm{km\,s^{-1}}$. The final BFs generated were smoothed with a Gaussian smoother of a 10 $\mathrm{km}\, \mathrm{s^{-1}}$ rolling window. We find that the BFs show wavelength-dependent variability. In Fig.\ref{fig:BF_ecl}, we show the BF variations observed in the spectra from secondary (Fig.\ref{fig:BF_ecl}; top panels) and primary (Fig.\ref{fig:BF_ecl}; bottom panels) eclipses. PP and OP in the plots denote pulsation phase and orbital phase, respectively. The time reference for the phase calculation is the same.

\section{Filtering Frequencies}

\label{appendix:pulsefilter}
\begin{figure*}
    \centering
    \includegraphics[width=0.3\textwidth]{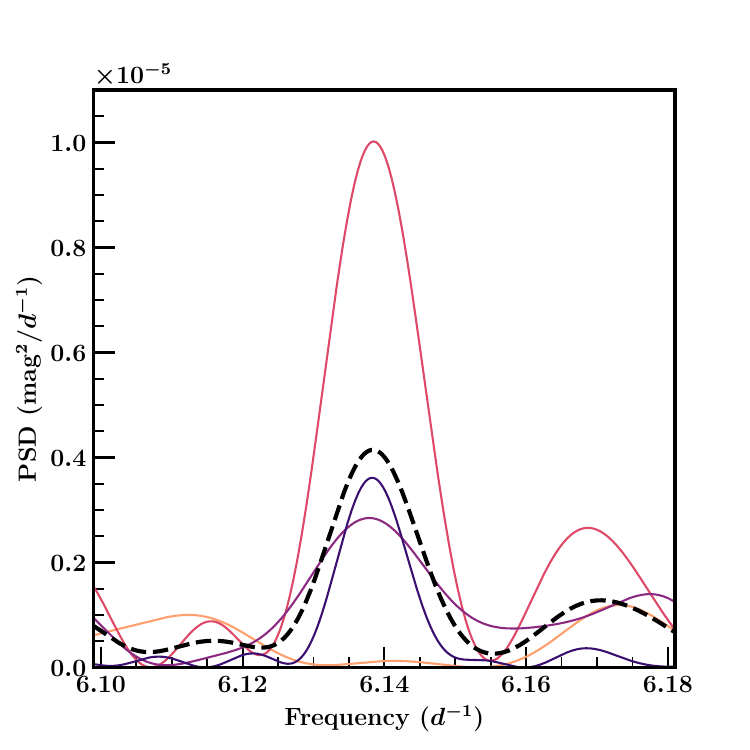}
        \includegraphics[width=0.3\textwidth]{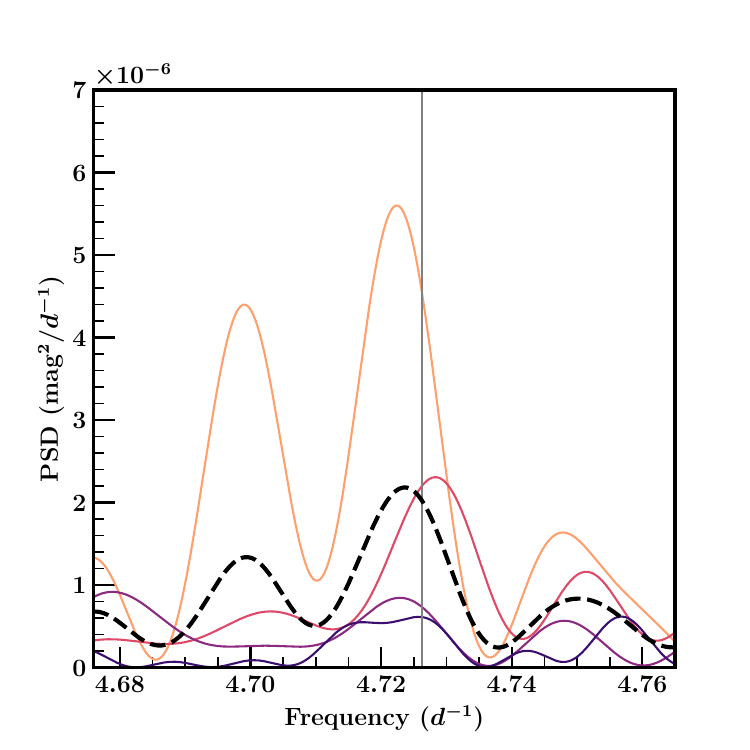}
            \includegraphics[width=0.3\textwidth]{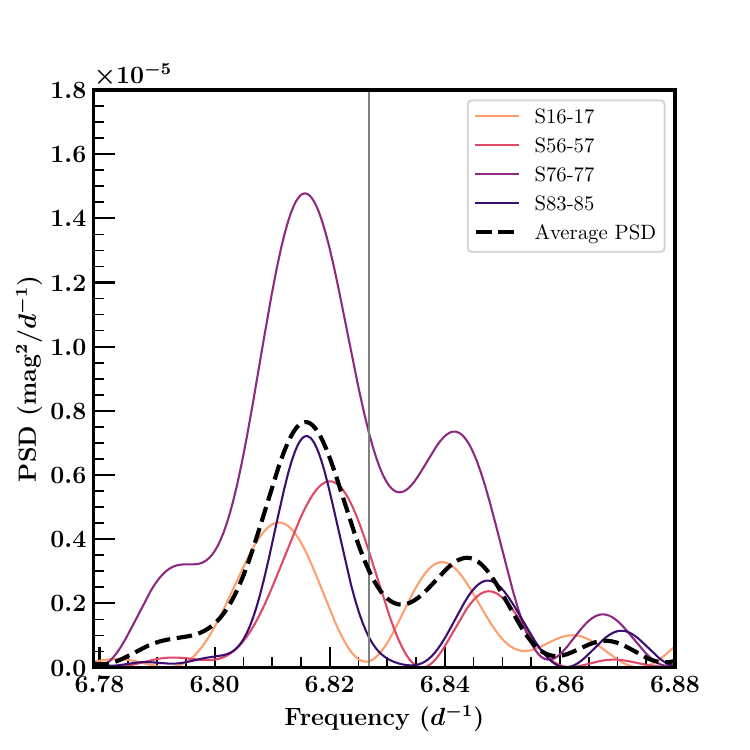} \\
    \includegraphics[width=0.3\textwidth]{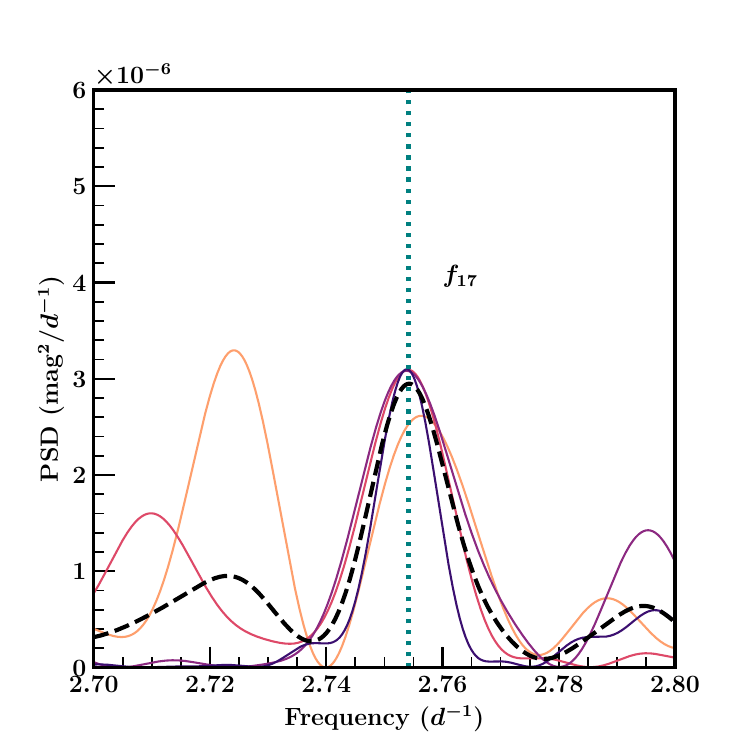}
        \includegraphics[width=0.3\textwidth]{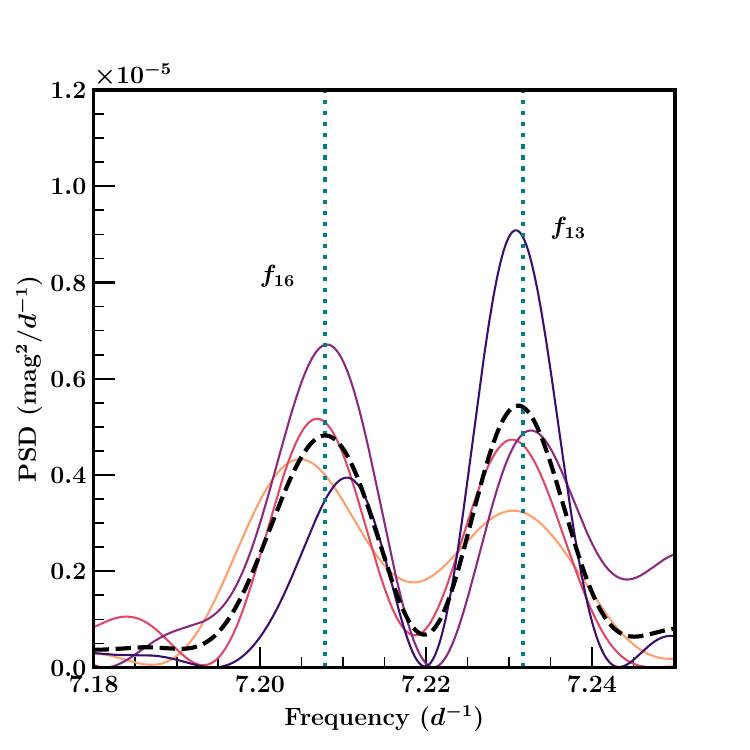}
            \includegraphics[width=0.3\textwidth]{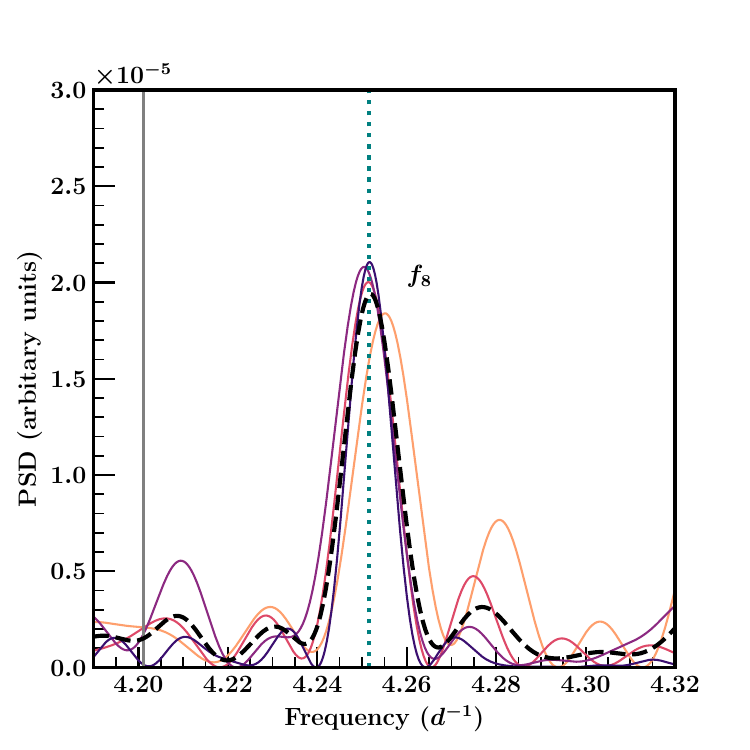}      \caption{PSD profiles for different sectoral sets and pulsations. The top panel shows the different frequencies that were found in the initial DFT search but rejected due to our stability criteria. The lower panel shows the accepted frequencies. The grey vertical lines represent the orbital harmonics and the teal dashed lines represent the position of the accepted frequencies.}
    \label{fig:pulse_PSDprofiles}     
\end{figure*}

We calculated power spectral density (PSD) for each set of consecutive sectors available (Fig.\ref{fig:pulse_PSDprofiles}) using the Lomb-Scargle periodogram. We visually inspected all the extracted frequencies that are higher than the $f_\mathrm{orb}$ and then rejected them if either: (i) they were not significant in all the light curve sets (Fig.\ref{fig:pulse_PSDprofiles}; top left), or (ii) they had S/N~$<5$ and were exact harmonics of $f_\mathrm{orb}$ (Fig.\ref{fig:pulse_PSDprofiles}; top middle), or (iii) they had extremely different PSD profiles in different sets (Fig.\ref{fig:pulse_PSDprofiles}; top right). We also rejected multiple frequencies detected in the DFT that were lower in amplitude than the central frequency but inside the central lobe of the window function of the PSD spectrum, and hence unresolved. The accepted frequencies are shown in the bottom panel of Fig.\ref{fig:pulse_PSDprofiles}.

\section{Tidally perturbed systems in the literature}
\label{app:rochelit}
 The list of tidally perturbed systems in the literature with measurements of masses and radii, is provided in Table \ref{tab:rochelit}. The Roche limit and orbital periods are also stated.  Tidally tilted systems are defined as systems where one pulsation axis is in the orbital plane, unlike tidally perturbed systems. Tidally trapped systems also have a tilted pulsation axis, but in addition, they have pulsations confined to a part of the stellar interior. For detailed definitions of the different kinds of tidally perturbed pulsators, we refer the reader to \citealt{SouthworthBowmanARAAreview}.

\begin{table*}
    \caption{ Tidally perturbed systems in the literature.}
    \footnotesize
    
    \begin{tabular}{lcccc}
    \hline
    System & \multicolumn{2}{c}{$R/R_\mathrm{Roche}$} & Period (d) & Source \\
    \hline
     & Pri. & Sec. & & \\
     \hline
\multicolumn{5}{l}{Tidally trapped} \\
\hline
TIC~63328020 &  0.98$^p$ & 0.99  & 1.10575 &  \cite{Rappaport_2021} \\
CO~Cam      &     0.73$^{p}$ & 0.74 &  1.27099 &  \cite{Kurtz_2020} \\
\hline
\multicolumn{5}{l}{Tidally tilted} \\
\hline
KIC~4851217   &  0.48 & 0.64$^{p}$ & 2.47039 &  \cite{Jennings_4851217} \\
TIC~435850195 &   0.66$^{p}$& 0.30 &  1.36719  &  \cite{TIC435850195_Jayaraman2024} \\
TIC~184743498  &  0.66$^{p}$ & 0.59 & 1.05324 &  \cite{TIC184743498_Zhang2024} \\
HD~265435    &    0.82$^{p}$ &  3.01 & 0.06882   & \cite{Jayaraman_2022} \\
RS~Cha        &   0.62 & 0.69$^{p}$ &  1.66988  &  \cite{Steindl2021} \\
{ TZ~Dra }     & { 0.65$^{p}$}  &   { 0.99} &  { 0.86603 }    &   { \cite{2022MNRAS.510.1413K} }\\
{ EL~CMi }     & { 0.71$^{p}$ } & { 0.99} &  { 1.05385} & {\cite{2025A&A...702A.104H}$^{1}$ } \\
\hline
\multicolumn{5}{l}{Tidally perturbed} \\
\hline
VV~Ori   &   0.83$^{p}$ & 0.60 & 1.48538 &  \cite{Southworth_2021} \\                
U~Gru     &       0.59$^{p}$ & 0.96 &  1.88050 &   \cite{Johnston_2023} \\
V453~Cyg   &      0.69$^{p}$ & 0.49 & 3.88976 &   \cite{Southworth_v453cyg} \\
\hline
\multicolumn{5}{l}{Perturbed g-mode} \\
\hline
KIC~3228863    &  0.87$^{p}$ & 0.99 & 0.73094 &  \cite{vanReeth_2023} \\
V456~Cyg       & 0.68 & 0.70$^{p}$ & 0.89119 &  \cite{vanReeth_2022} \\
\hline 
    \end{tabular} \\
     $^{p}$ perturbed component $^1$ from \phoebe\, solution  

    \label{tab:rochelit}
\end{table*}

\section{Astrometry simulations of systems with known 3D geometry}
\label{appendix:chtruwe}
\begin{figure}
    \centering
    \includegraphics[width=0.9\columnwidth]{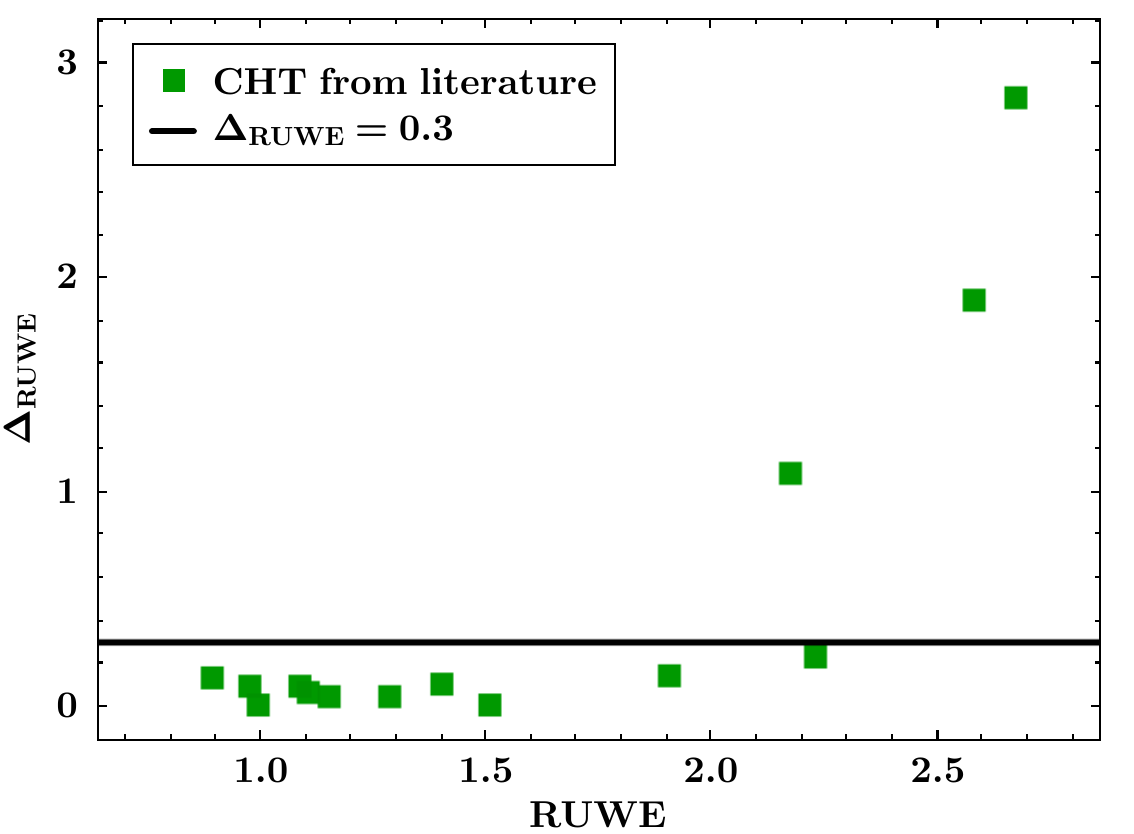}
    \caption{Comparison of $\Delta_\mathrm{RUWE}$ for different compact hierarchical triples with measurements of all orbital parameters. $\Delta_\mathrm{RUWE}$ is calculated by taking the absolute difference of RUWE calculated using a \textsc{gaiamock} model and the actual RUWE value observed with GAIA.}
    \label{fig:chtruwe}
\end{figure}

We compiled a list of compact hierarchical triples (CHT), which have precisely measured parameters that define a system's 3D geometry. From \cite{Borkovits_2020,Borkovits_2022,Moharana_ST3,Moharana_ST3n2}, and \cite{Rappaport_2024}, we get all orbital parameters, including $i_{2}$ and $\Omega_2$. We then calculated theoretical RUWE using \textsc{gaiamock} and compared the deviation ($\Delta_\mathrm{RUWE}$) from observed RUWE values. We find that most of the systems have consistent observed and simulated RUWE, within an error of 0.3 (Fig.\ref{fig:chtruwe}). The systems that have a large discrepancy with the simulations have RUWE greater than 2. This shows that models with $\Delta_\mathrm{RUWE}$ less than 0.3 are acceptable solutions for \vcep.  

\section{Eclipse timings}\label{app:ecl_times}

In Table~\ref{tab:ecl_times}, we provide the times of primary and secondary minima of \vcep{} derived from the \textit{TESS} light curve using the procedure described in Section~\ref{sec:etv}.

\begin{table}
 \centering
 \caption{
 Times of minima of the primary and secondary eclipses from the \textit{TESS} light curve of \vcep{}. The full table is available in machine-readable format in the online supplementary material.}
 \label{tab:ecl_times}
 {
        \renewcommand{\arraystretch}{1.0}
 \begin{tabular}{@{}lcc@{}}
            \hline
            \hline
            Time                & Cycle & 1$\sigma$ error \\ \relax
            BJD$-2457000$       & no.   & (d)             \\
            \hline
            1740.34139          & 0.0   & 0.00018         \\
            1742.26266          & 0.5   & 0.00042         \\
            1744.15017          & 1.0   & 0.00017         \\
            1746.07244          & 1.5   & 0.00049         \\
            1747.95882          & 2.0   & 0.00017         \\
            1751.76608          & 3.0   & 0.00023         \\
            1753.68944          & 3.5   & 0.00039         \\
            1755.57589          & 4.0   & 0.00017         \\
            1757.49808          & 4.5   & 0.00038         \\
            1759.38381          & 5.0   & 0.00017         \\
            \hline
        \end{tabular}
        } \\
        \textit{Notes}. Half-integer cycle numbers refer to secondary eclipses. 
\end{table}

\section{Period change measurements}

 The period change measurements for the dominant period are given in Table.\ref{tab:pcs}.

\begin{table}
\centering
\caption{Period change measurements used in this work.}
\label{tab:pcs}
        \renewcommand{\arraystretch}{1.0}
 \begin{tabular}{@{}lcc@{}}
\hline
BJD-2457000 & Period Changes (d) & Errors (d) \\
\hline
1745.242110 & 0.00732 & 0.00014 \\
1763.289777 & 0.00621 & 0.00014 \\
1782.019033 & 0.00583 & 0.00014 \\
1956.072593 & -0.00313 & 0.00015 \\
1975.792280 & -0.00405 & 0.00011 \\
2830.179857 & -0.01015 & 0.00013 \\
2840.161379 & -0.00968 & 0.00014 \\
2850.594956 & -0.00996 & 0.00012 \\
2864.472345 & -0.00883 & 0.00013 \\
2877.217883 & -0.00829 & 0.00012 \\
3371.913141 & 0.00130 & 0.00014 \\
3380.496760 & 0.00194 & 0.00012 \\
3389.084658 & 0.00178 & 0.00014 \\
3400.321187 & 0.00169 & 0.00015 \\
3416.962323 & 0.00212 & 0.00016 \\
3566.970956 & 0.00224 & 0.00010 \\
3581.906368 & 0.00248 & 0.00009 \\
3598.187738 & 0.00296 & 0.00011 \\
3613.802411 & 0.00243 & 0.00010 \\
3628.743467 & 0.00250 & 0.00010 \\
\hline
\end{tabular}
\end{table}
\bsp 
\label{lastpage}
\end{document}